\documentclass{emulateapj}
\usepackage{amsmath}
\usepackage{color}
\usepackage{ulem}
\usepackage{upgreek}
\usepackage{subfigure}
\usepackage{mathtools}
\usepackage{lipsum}
\usepackage{bm}

\shorttitle{Mira Variable Stars in the LAMOST DR4 Data}
\shortauthors{Y.-H. Yao et al.}
\slugcomment{Submitted for publication in The Astrophysical Journal Supplement Series}
\begin{document}

\title{Mira Variable Stars From LAMOST DR4 Data: 
	\\Emission Features, Temperature Types, and Candidate Selection}
\author{Yuhan Yao$^{1}$,
Chao Liu$^{2}$,
Licai Deng$^{2}$,
Richard~de~Grijs$^{1,3}$, and
Noriyuki Matsunaga$^{4}$}
\affil{$^1$Kavli Institute for Astronomy \& Astrophysics and
  Department of Astronomy, Peking University, Yi He Yuan Lu 5, Hai
  Dian District, Beijing 100871, China\\
$^2$Key Laboratory for Optical Astronomy, National Astronomical
  Observatories, Chinese Academy of Sciences, 20A Datun Road, Chaoyang
  District, Beijing 100012, China \\
$^3$International Space Science Institute--Beijing, 1 Nanertiao, Hai
  Dian District, Beijing 100190, China \\
$^4$Department of Astronomy, The University of Tokyo, 7-3-1 Hongo,
  Bunkyo-ku, Tokyo 113-0033, Japan}
  
\begin{abstract}
Based on an extensive spectral study of a photometrically confirmed
sample of Mira variables, we find a relationship between relative
Balmer emission-line strength and spectral temperature of O-rich Mira
stars. The $F_{\rm H\delta}/F_{\rm H\gamma}$ flux ratio increases from
less than unity to five as stars cool down from M0 to M10, which is
likely driven by increasing TiO absorption above the deepest
shock-emitting regions. We also discuss the relationship between the
equivalent widths of the Balmer emission lines and the photometric
luminosity phase of our Mira sample stars. Using our 291 Mira spectra
as templates for reference, 191 Mira candidates are newly identified
from the LAMOST DR4 catalog. We summarize the criteria adopted to
select Mira candidates based on emission-line indices and molecular
absorption bands. This enlarged spectral sample of Mira variables has
the potential to contribute significantly to our knowledge of the
optical properties of Mira stars and will facilitate further studies
of these late-type, long-period variables.
\end{abstract}


\section{Introduction} \label{sec:intro}

Mira stars are a class of late-type, long period variables (LPVs) that
coincide with the coolest, most luminous part of the asymptotic giant
branch (AGB). The AGB is the final stellar evolutionary phase of low-
to intermediate-mass (1--8 $M_\odot$) stars before the envelope
ejection phase \citep{2003agbs.conf.....H}. Miras are, by definition,
AGB variable stars exhibiting large luminosity amplitudes
(\textgreater 2.5 mag in the $V$ band) with periods in excess of 80
days.

Just like other AGB giants, Miras can be divided into M-, S-, and
carbon N-type stars. The differences among these groups are defined in
terms of the relative abundances of oxygen and carbon in their
photospheres. In M-type stars, a small amount of carbon in the stellar
atmosphere is largely consumed by carbon monoxide (CO), so there is
surplus oxygen left to form molecules such as TiO. During the
thermally pulsating AGB (TP-AGB) phase, s-process elements (e.g., ZrO)
and carbon from deep inside the star can be dredged up to the surface
in stars with sufficiently large initial masses. Stars containing both
TiO and ZrO are known as MS stars. As a star evolves up the AGB, the
dredge-up continues to enrich the stellar envelope, so more carbon
soaks up all excess oxygen into CO, leading to pure S or SC
stars. Further enrichment brings up plenty of carbon to form C$_2$,
CN, and/or CH molecules, which dominate the spectra of carbon N-type
stars \citep{2003agbs.conf.....H, 2013ApJ...765...12G,
  2017A&A...601A..10V}. As soon as the C/O ratio increases to above
unity, the difference in chemical composition leads to a significant
decrease in the effective temperature, $T_{\rm eff}$
\citep{2008A&A...482..883M}.

Based on extensive observations of Mira stars since the 1940s, it has
been found that one of the major spectral characteristics of these
stars resides in their high-excitation emission lines. These lines
vary over the pulsation cycle, but they are strongest around maximum
light and weakest after minimum luminosity
\citep{2009ssc..book.....G}.  Shock-heating of the atmosphere is
verifiably at the basis of these variations
\citep{1959ApJ...130..570D, 1961SvA.....5..192G, 1976ApJ...205..172W,
  2004A&A...420..423F}.

For O-rich Miras, the emission-line flux of H$\delta$ is greater than
that of H$\gamma$, which, in turn, is greater than that of H$\beta$;
H$\alpha$ is weakest among the four lines in Balmer series. This
so-called `Balmer increment' is not found in C-rich stars, which show
the opposite behavior \citep{1940slpv.book.....M}. This tendency for
O-rich Miras was initially attributed to obscuration by overlying
absorption \citep{1945PASP...57..178M,
  1947ApJ...106..288J}. \citet{1992ASPC...26..558L} subsequently
proposed that non-local thermal equilibrium (NLTE) radiative transfer
on its own can also give rise to the phenomenon. However, their
interpretation, based on \citet{1988ApJ...329..299B}'s models, assumed
the existence of an extended chromosphere, which does not exist in
real Mira variables. Instead, it is only the innermost shock region
that produces the observable Balmer-line emission
\citep{1984ApJ...286..337F, 2003A&A...400..319R}.

Apart from the hydrogen Balmer series, many metallic emission lines,
such as Mg{\sc i}, Mn{\sc i}, Si{\sc i}, Fe{\sc i}, and Fe{\sc ii}, as
well as forbidden [Fe{\sc ii}] emission lines, are also reported for
M-type Mira stars. However, metallic emission lines appear late in the
pulsation cycle, and vary independently of the hydrogen
lines\citep{1954ApJS....1...39J, 2001A&A...369.1027R,
  2010BASI...38....1G}.

Mira variables have long been used as tracers of stellar populations,
because they are visible beyond the Local Group
\citep{2004A&A...413..903R}. Although Mira period--luminosity (PL)
relations exhibit an intrinsic dispersion in the optical \citep{de
  Grijs 2011}, it has been found that such dispersions are smaller at
near- and mid-infrared (IR) wavelengths for Miras with thin dust
shells and periods of less than 400 days \citep{2008MNRAS.386..313W,
  2009MNRAS.399.1709M, 2013IAUS..289..209W}. Thus, their PL relations
can be used to trace the structure of the Milky Way (MW) galaxy and
beyond \citep{2016MNRAS.455.2216C, 2017ApJ...836..218L}. Moreover, SiO
masers from Mira stars have proved powerful tools for investigating
stellar motions in optically obscured regions of the MW
\citep{2006PASJ...58..529F, 2007PASJ...59..559D, 2008ASSP....4...33D,
  2010PASJ...62..525D, 2012IAUS..287..265D}. Stepping into the {\sl
  Gaia} era, a first result from {\sl Gaia} Data Release 1 (DR1) used
Miras to probe the outer regions of the Large Magellanic Cloud
\citep{2017MNRAS.467.2636D}. Therefore, increasing the number of
either type of Mira stars would be very helpful for similar future
studies.

Generally, the study of Mira stars involves both long-term IR
photometric and carefully planned spectroscopic observations.
Therefore, for years optical spectroscopy of Mira stars has focused on
well-known targets. Thanks to the wide-field Large sky Area
Multi-Object fiber Spectroscopic Telescope (LAMOST) survey, we now
have a chance to perform a novel, comprehensive study of Mira
variables, investigate their overall optical spectral properties, and
provide insights into the physical mechanisms driving these late-type
stars.

This paper is organized as follows. A brief introduction to the LAMOST
survey and the sources of our photometrically confirmed Mira sample
are given in Section \ref{sec:source}. In Section \ref{sec:templates},
we provide a detailed description of our template Mira spectra,
discuss new observational phenomena, and suggest possible stellar
processes for interpretation. In Section \ref{sec:identification}, we
outline the method used to search for Mira candidates in the LAMOST
catalog. A discussion and conclusions of this work are given in
Section \ref{sec:conclusion}.


\section{Data}
\label{sec:source}
\subsection{LAMOST}\label{subsec:lamost}

LAMOST is a 4 m-diameter reflective Schmidt telescope equipped with
4000 fibers in its focal plane, allowing it to obtain spectra covering
the wavelength range from 380 to 900 nm at a spectral resolution of
$R$ = 1800 \citep{2012RAA....12..723Z, 2012RAA....12.1197C}.  The
five-year General Survey of LAMOST (2012--2017) is mainly a Galactic
stellar survey. As a major component, the `LAMOST Experiment for
Galactic Understanding and Exploration' (LEGUE) survey has been
designed to target a uniform and (statistically) complete sample of
all stellar populations \citep{2012RAA....12..735D}. Therefore, Miras,
as evolved AGB stars, are included in its observations.

By the end of March 2016, the LAMOST DR4 catalog had accumulated
7,681,185 spectra, of which 6,898,298 were of stars. These 1D spectra
have been processed by the LAMOST data reduction system. Basic
reduction steps such as de-biasing, flat-fielding, fiber tracing, sky
subtraction, and wavelength calibration are included in its 2D
pipeline. Absolute flux calibration has not been done. However,
relative flux calibration was performed (in terms of the flux per unit
wavelength) by selecting stars with high-quality spectra as standard
stars. Using a cross-correlation method, the 1D pipeline assigns a
spectral type and redshift to each 1D spectrum
\citep{2012RAA....12.1243L, 2015RAA....15.1095L}.

In this paper, we measure the spectral indices of LAMOST spectra. The
relevant wavelength ranges of the indices and their shoulder
bandpass(es) are included in Table \ref{tab:definition}. The
definitions of the line indices, in terms of their integrated flux
($F$; {Section \ref{subsubsec:O-rich}}), equivalent width (EW;
Sections \ref{subsec:sample_emission}, \ref{subsec:Fe/Mg emission},
and \ref{subsec:Balmer_selection}), and the band indices ($R_{\rm
  ind}$; Section \ref{subsec:giants}) will be included when they are
first used in the text.

\begin{deluxetable}{c|c|c}
\tablecaption{Line Index Definitions \label{tab:definition}}
\tablehead{\colhead{Name} & \colhead{Index Bandpass (\AA)} & \colhead{Shoulder Bandpass(es) (\AA)}}\\
\startdata
     H$\delta$          & 4096--4109                  & 4085--4094, 4112--4122\\
     Fe4202             & 4197--4208                  & 4183--4192, 4208--4217\\
     Fe4308             & 4304--4314                  & 4291--4302, 4315--4323\\
     H$\gamma$      & 4332--4347                  & 4320--4332, 4349--4352\\
     Fe4376             & 4370--4382                  & 4361--4370, 4383--4392\\
     Mg4571             & 4567--4576                  & 4556--4566, 4578--4584\\
     H$\beta$           & 4857--4868                  & 4845--4855, 4875--4880\\
     H$\alpha$          & 6548--6578                  & 6505--6540, 6580--6610\\
     CaH2                & 6814--6846                  & 7042--7046\\
     CaH3                & 6960--6990                  & 7042--7046\\
     TiO5                  & 7126--7135                  & 7042--7046
\enddata
\end{deluxetable}

\subsection{A Photometrically Confirmed Sample}
\begin{figure*}[ht!]
\center
\centering
\includegraphics [scale=0.70] {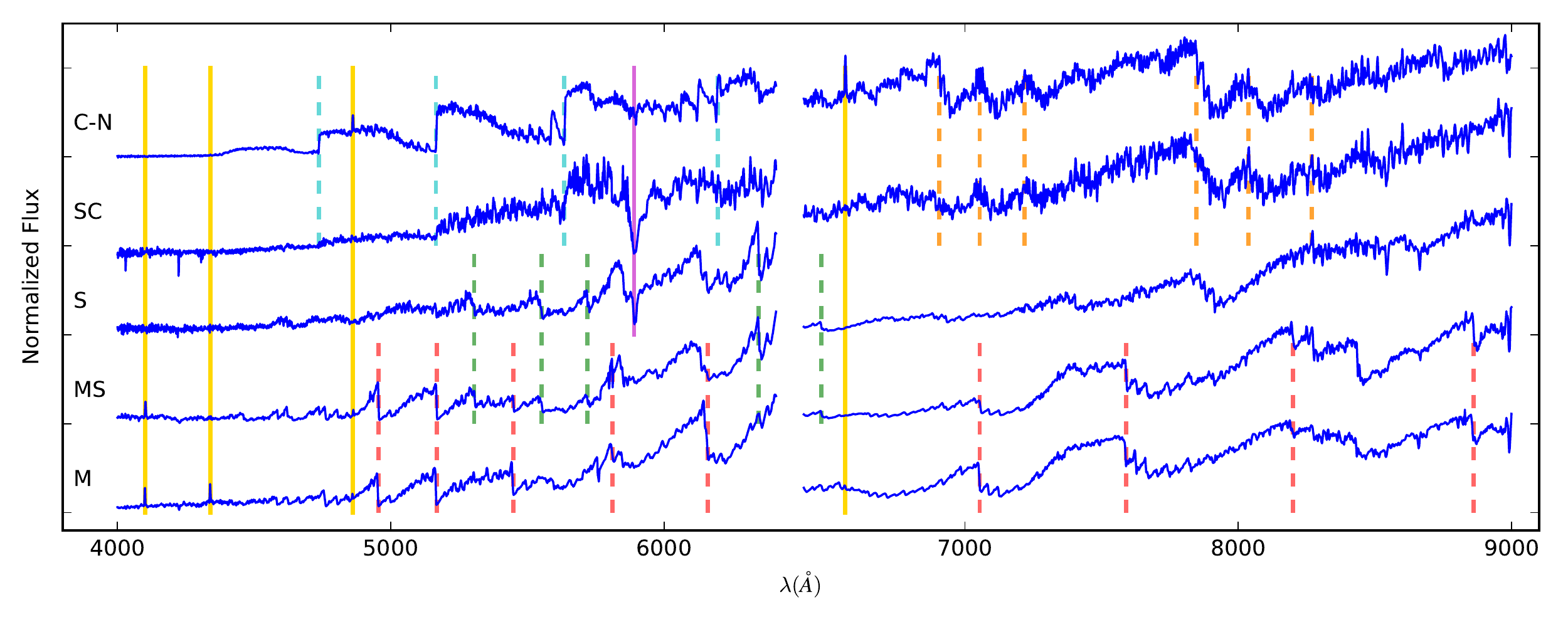}
\caption{(bottom to top) The M $\rightarrow$ MS $\rightarrow$ S
  $\rightarrow$ SC $\rightarrow$ C spectral sequence of LAMOST Mira
  spectra. Each spectrum is normalized at both 6315\AA \ and
  8750\AA. Solid lines indicate emission-line wavelengths (yellow:
  Balmer series; purple: Na{\sc i} D doublet). Dashed lines indicate
  band heads of molecular absorption bands (red: TiO; green: ZrO;
  cyan: C$_2$; orange: CN). \label{fig:sequence}}
\end{figure*}

We compiled a photometrically confirmed sample of Mira variables from
the Kiso Wide-Field Camera (KWFC) Intensive Survey of the Galactic
Plane (KISOGP) \citep{2017arXiv170508567M}, the American Association
of Variable Star Observers (AAVSO) International Database Variable
Star Index \citep[VSX;][version 2017-05-02; we selected stars of
  variability type `M']{2006SASS...25...47W}, and the SIMBAD
Astronomical Database.

The KISOGP survey uses the KWFC mounted on the 105 cm Schmidt
telescope at Kiso observatory, Japan. Having obtained $I$-band
photometric measurements of $m_I = 9.5$--16.5 mag, from 2012 to 2015,
this survey identified more than 700 Mira stars in the northern
Galactic disk with periods between 100 and 600 days, of which roughly
90\% were newly found.

We first cross-matched the KISOGP and VSX Miras with the LAMOST DR4
catalog. Spectra that suffer from serious contamination by sky lines,
low signal-to-noise ratios (SNRs), flux saturation, or which display
characteristics of a binary spectrum were excluded from the Mira
sample. This resulted in 19 and 238 spectra for, respectively, KISOGP
and VSX targets. (Eight spectra were in common between the VSX and
KISOGP catalogs; they were only included among the 19 spectra with
KISOGP designations.)

For KISOGP targets, the period and epoch (date of light maximum) were
calculated from the KISOGP light curves using generalized Lomb Scargle
(GLS) fitting \citep{2009A&A...496..577Z}, which is a common approach
to estimate the sinusoidal primary frequency. The phases of each
spectrum were calculated using
\begin{equation}
\phi = \frac{{\rm (observed \ date} - epoch) \ {\rm mod} \ period}{period} \label{eq:phase}
\end{equation}

As regards VSX targets, if period or epoch information was available
in the VSX catalog, we estimated the phase using Eq. (\ref{eq:phase}),
and adopted the resulting value if
\begin{equation}
\frac{{\rm observed \ date} - epoch}{period} < 18 \label{eq:18}
\end{equation}
to ensure accuracy. For targets without such information, we searched
for their light curves in the Northern Sky Variability
Survey\footnote{http://www.skydot.lanl.gov/nsvs/nsvs.php}
\citep[NSVS]{2004AJ....127.2436W} and the All Sky Automated
Survey\footnote{http://www.astrouw.edu.pl/asas/?page=aasc}
\citep[ASVS]{2014CoSka..43..523P}. If available, we fitted the light
curve and repeated the steps of Eqs (\ref{eq:phase}) and
(\ref{eq:18}).

Finally, we cross-matched the DR4 catalog with the SIMBAD database,
because some Mira stars have a SIMBAD designation but are not included
in the VSX or KISOGP catalogs. This procedure resulted in the
identification of another 34 Mira spectra,\footnote{24 of them are
  also tabulated in the VSX, 22 of which as semi-regular variables and
  the other two are marked as long-period variables.} resulting in a
total number of 291 Mira spectra.


\section{Mira Spectra}
\label{sec:templates}
\subsection{Overview: the Spectral Sequence} 

Based on the TiO, ZrO, C$_2$ and CN molecular bands, we broadly
divided the 291 Mira spectra into O-rich (258) and C-rich (33)
Miras. Among the C-rich spectra, six are of type SC, which show no
TiO, weak or no ZrO/C$_2$, and very strong D lines, as defined in the
Boeshaar--Keenan System \citep{1980ApJS...43..379K}. Among the O-rich
spectra, about 20 are of types MS or S, which show prominent ZrO
features. We show one spectrum of each type in Figure
\ref{fig:sequence}. The Balmer {\it decrement} in C-rich Miras
($F_{{\rm H}\delta} < F_{{\rm H}\gamma} < F_{{\rm H}\beta} < F_{{\rm
    H}\alpha}$) and the Balmer {\it increment} in O-rich Miras are
also clearly seen.

Molecular dissociation calculations \citep{1976A&A....48..219S} have
demonstrated that the sequence M $\rightarrow$ MS $\rightarrow$ S
$\rightarrow$ SC $\rightarrow$ C represents an increase in the ratio
of carbon to oxygen; this ratio passes through a value of unity in SC
stars \citep{2009ssc..book.....G}. The sample of LAMOST Mira spectra
covers the entire evolutionary sequence of Mira variables in the
TP-AGB phase.

\subsection{O-rich}
\subsubsection{Balmer Emission in Subtypes}\label{subsubsec:O-rich}

From among the 258 O-rich spectra, we selected 104 with observable
Balmer emission(s), and assigned a temperature type to each of them by
comparison with the LAMOST M0--M6 giant templates from
\citet[hereafter Z15]{2015yCatp040001501Z}, as well as the M0--M10
intrinsic templates derived by \citet{1994A&AS..105..311F}. From a
careful comparison of these spectral templates, it follows that the
Balmer increment relation only holds for late-type O-rich Miras
(M5--M10). As for the early-type spectra, we see a Balmer {\it
  decrement} for M1 and M2, but H$\alpha$ is weak in the only M0
spectrum in our sample. We also observe nearly equivalent strength for
the Balmer lines in the case of M3. For spectra later than M4, the
strengths of H$\alpha$ and H$\beta$ decrease and H$\delta$ becomes the
strongest line.

We use the flux ratio of H$\delta$ and H$\gamma$ to quantify this
relationship, because they are visible for the longest duration. We
define the integrated line flux as
 \begin{equation}
      F = \int \left( f_{\rm line}(\lambda) - f_{\rm
          cont}(\lambda) \right) {\rm d}\lambda, \label{eq:flux}
    \end{equation} 
where $f_{\rm line}(\lambda)$ is the flux of the spectral line in the
relevant index bandpass (see Table \ref{tab:definition}) and $f_{\rm
  cont}(\lambda)$ is the pseudo-continuum estimated through linear
interpolation in the shoulder bandpass(es).

\begin{figure}
\center
\centering
\includegraphics [scale=0.7] {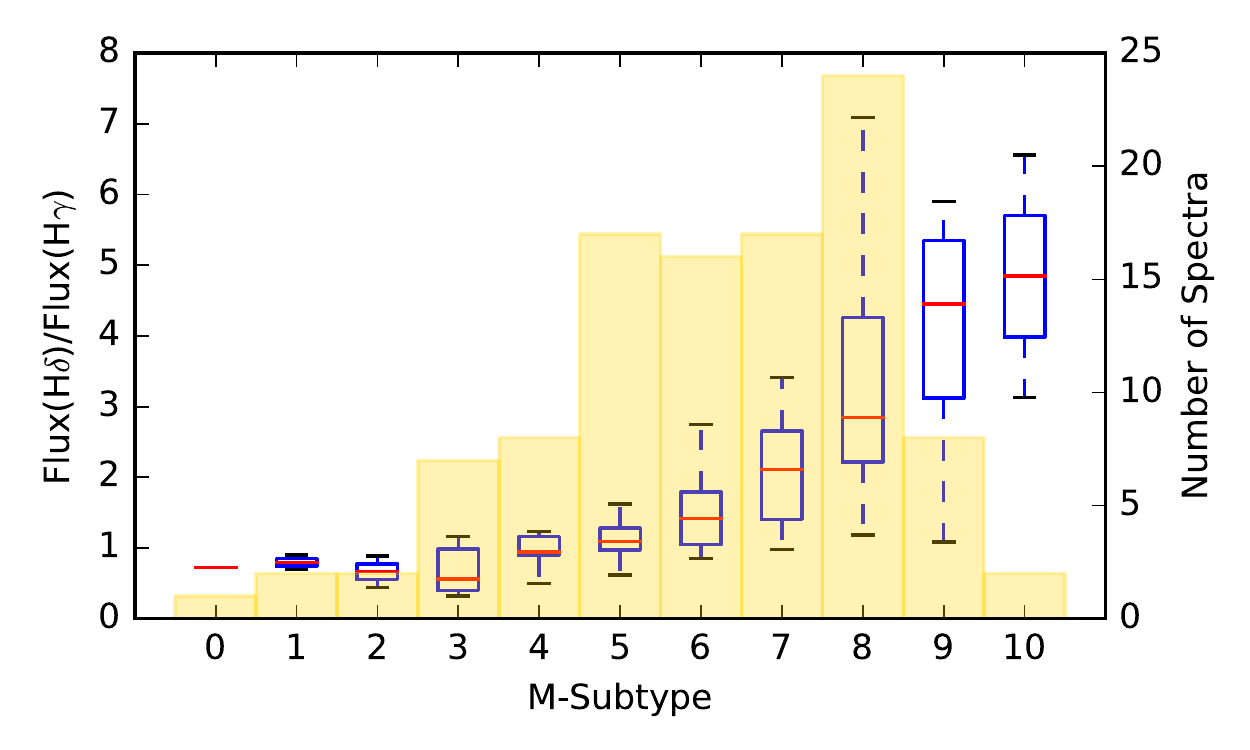}
\caption{Box-and-whisker plot of the $F_{\rm H\delta}/F_{\rm H\gamma}$
  ratio for each subtype of O-rich Mira templates. The box extends
  from the lower to upper quartile values, with a red line at the
  median. The whiskers extend from the box to show the range of the
  data. The yellow histogram indicates the number of spectra used for
  each subtype.\\
\label{fig:ratio}}
\end{figure}

Box-and-whisker plots of $F_{\rm H\delta}/F_{\rm H\gamma}$ for each
subtype are shown in Figure \ref{fig:ratio}, with the background
yellow histogram indicating the number of spectra used for each
subtype. Our sample is dominated by M5--M8. From M0 to M3 the ratio
does not change much; it even decreases somewhat. The median inverts
from `$<$1' to `$>$1' between M4 and M5. For spectra later than M5,
the ratio increases. More M0--M4 and M9--M10 templates are needed to
render this relation more accurate. From Figure \ref{fig:ratio}, we
are aware of the large intrinsic variance within each type, so this
relationship cannot be used individually to assign spectral types to
optical spectra.

Early observations by \citet{1945PASP...57..178M} of bright Mira stars
revealed that the low intensities of H$\alpha$, H$\beta$, and
H$\gamma$ are mainly owing to absorption by TiO molecular bands.
These overlapping absorptions originate above the deepest shock layer,
and they can thus effectively block radiation from internal
atmospheric layers where the Balmer lines are formed
\citep{2016IBVS.6183....1S}. Our result agrees well with this
scenario: as an AGB star cools down from M0 to M10, the TiO bands
strengthen gradually, so more flux is absorbed by the lower-order
Balmer lines, leading to an increase in $F_{\rm H\delta}/F_{\rm
  H\gamma}$. Note that narrow portions of H$\delta$ and H$\gamma$ are
sometimes absorbed by superimposed metal lines of iron, vanadium,
yttrium, and indium \citep{1941ApJ....93...11A, 1947ApJ...106..288J},
but this effect has little influence on the overall integrated flux of
low-resolution spectra.

Also note that all five M0--M2 spectra in our O-rich sample have
strong Balmer lines, regardless of our knowledge that the emission
strengths are strongly correlated with luminosity phase.

\subsubsection{Metallic Emission}\label{subsubsec:metallic emission}

\begin{figure*}
\center
\begin{tabular}{c}
\includegraphics [scale=0.58] {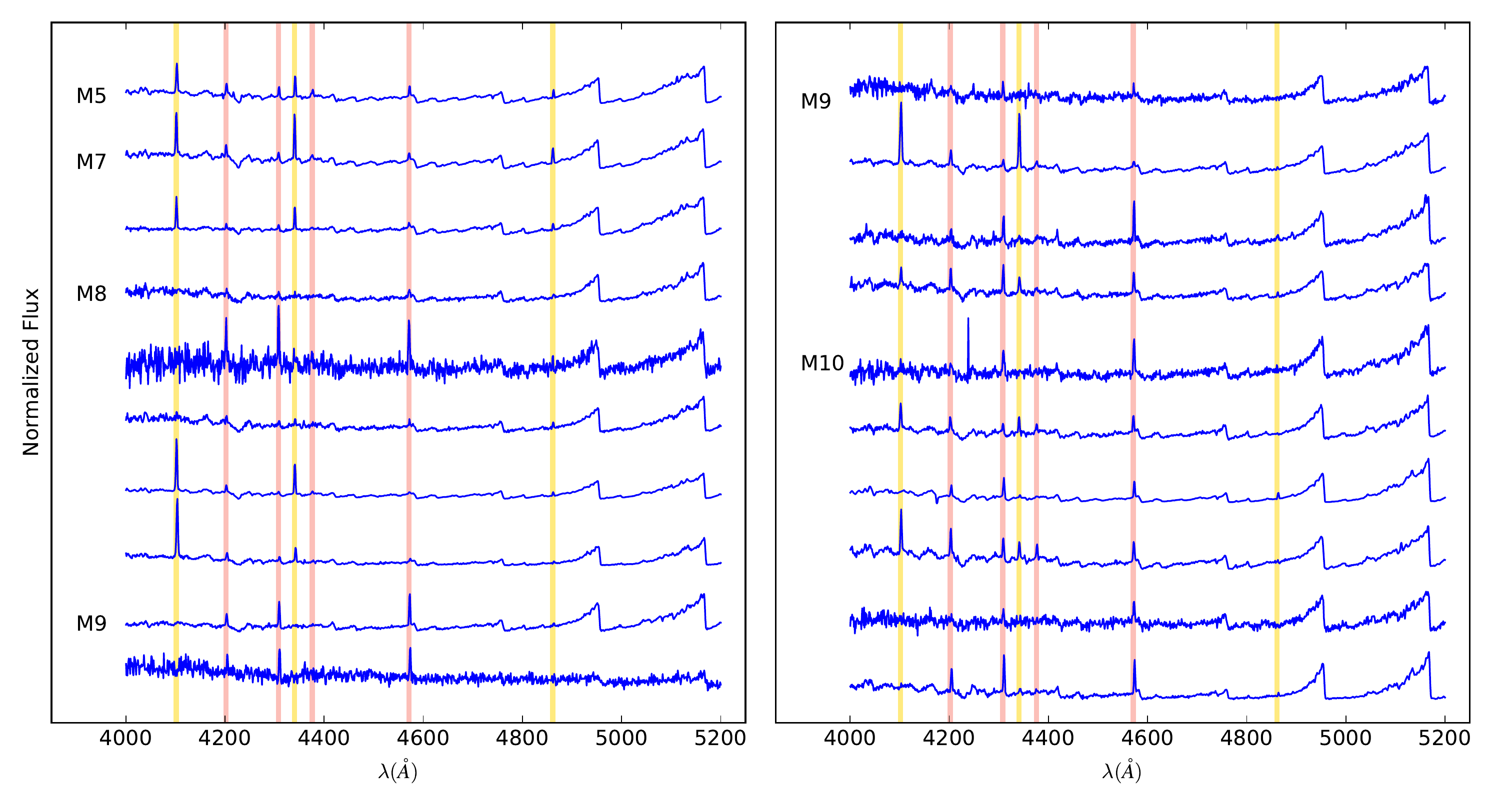}\\
\end{tabular}
\caption{Twenty O-rich templates with metallic emission normalized at
  5175 \AA. Colored lines indicate the emission wavelength (yellow for
  Balmer lines; pink for metallic lines). \label{fig:femg} }
\end{figure*}

Among the 258 O-rich sample stars, we can distinguish four metallic
lines in 20 spectra, as shown in Figure \ref{fig:femg}. The emission
lines are Fe {\sc i} $\lambda\lambda$4202, 4308, 4376 and Mg {\sc i}
4571, all of which are well-known emission features of Mira stars
\citep{2009ssc..book.....G}. We note that 17 of the 20 spectra are of
types M8--M10, with M5 being the earliest type. This is a strong
indication that these four emission lines are mostly developed in
late-type O-rich Miras.

\subsection{C-rich Spectra}\label{study_c}

All of our C-rich spectra are included in the Appendix (see Figure
\ref{fig:cs2}). The small number of C-rich sample stars, as well as
the difficulty in finding a clean control sample of non-variable
carbon N-type spectra, prevents us from carrying out a detailed study
similar to what we did for our O-rich sample.

However, there are still two features of interest. First, more than
half of the 33 spectra exhibit very strong H$\alpha$ emission, much
stronger than the weak emission shown in \citet[][the Bambaum Atlas
  includes a sample of non-variable carbon N-type
  stars]{1996ApJS..105..419B}. Second, except for two SC-type stars,
none of our C-rich sample spectra show the Ca{\sc ii}
$\lambda\lambda$8498,8542,8662 triplet.

\subsection{Emission as A Function of Phase}\label{subsec:sample_emission}
\begin{figure*}
\center
\centering
\includegraphics [scale=0.48] {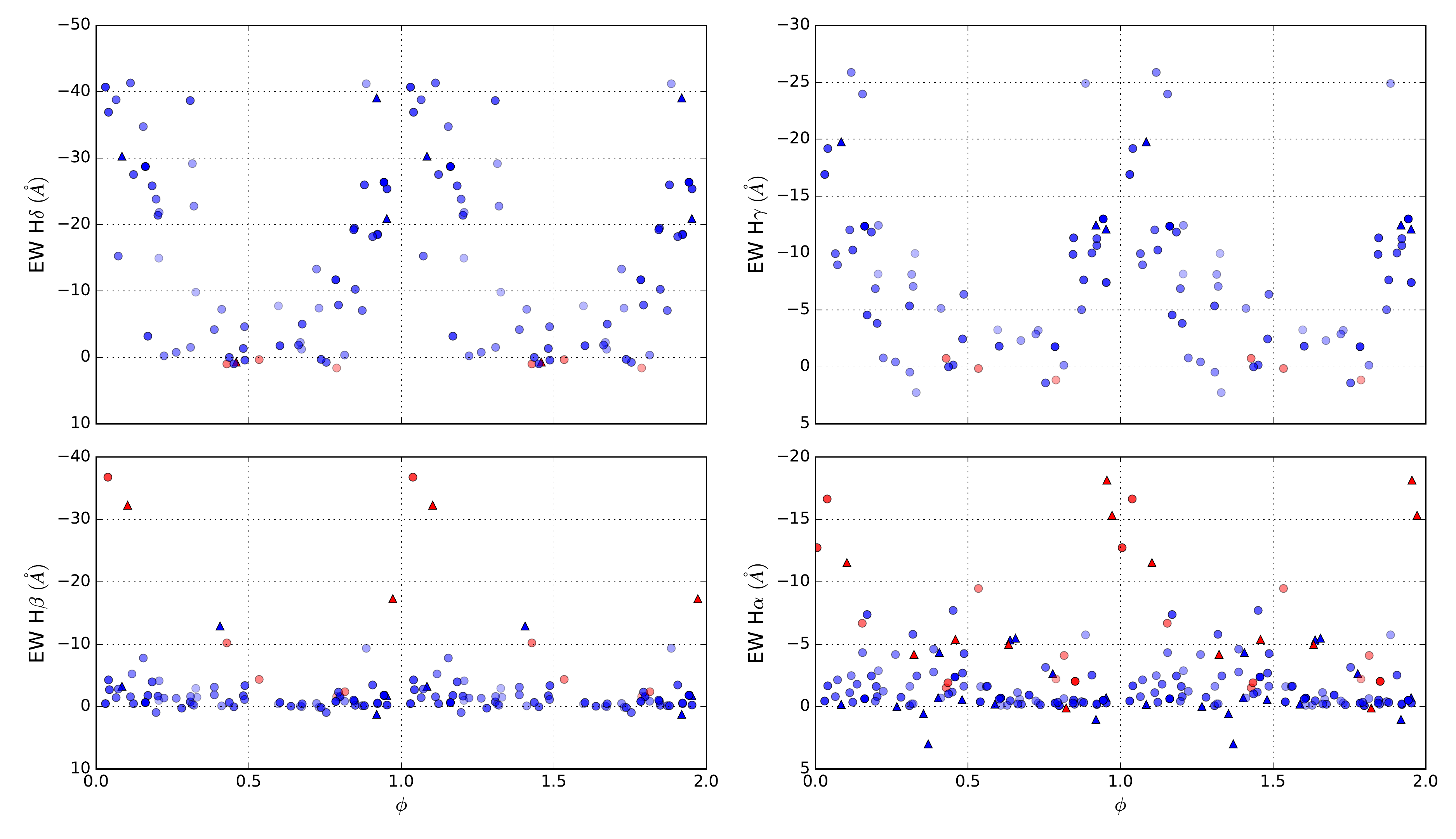}
\caption{Balmer emission strength (EW) as a function of pulsation
  phase ($\phi$). O- and C-rich spectra are plotted in blue and red,
  respectively. Triangles and circles are KISOGP and VSX targets,
  respectively. For phases calculated from VSX data and NSVS/ASVS
  light curves, we use variations in transparency to indicate the
  accuracy (the deeper the color, the smaller the value of
  Eq. \ref{eq:18}). \label{fig:elements}}
\end{figure*}

We will next investigate the relationship between the Balmer emission
strength and the corresponding luminosity phase, $\phi$. Use of the
flux-based definition of the line index, Eq. (\ref{eq:flux}), is most
appropriate when comparing fluxes of different lines based on a single
spectrum, but here we need to characterize the emission strengths for
all emission spectra in our sample. Hence, we use the EW
\citep{1994ApJS...94..687W} as an indication of line strength. The EW
is defined as
  \begin{equation}
      {\rm EW} = \int \left(1 - \frac{f_{\rm line}(\lambda)}{f_{\rm
          cont}(\lambda)}\right) {\rm d}\lambda ,\label{eq:EW}
  \end{equation} 
where $f_{\rm line}(\lambda)$ and $f_{\rm cont}(\lambda)$ have the
same meaning as in Eq. (\ref{eq:flux}).

For spectra with low SNRs, fluxes in the shoulder bandpass(es) tend to
be noisy, and thus the pseudo-continuum estimated through linear
interpolation does not always represent the continuum level very
well. Hence, we define a local SNR for every line index to quantify
the significance of the measured EW. Using the concept of the inverse
variance,
    \begin{equation}
      {\it SNR}_{\rm local}=\frac{1}{{\rm Var}(f_{\rm
          line}(\lambda)-f_{\rm cont}(\lambda))} ,\label{eq:SNR}
    \end{equation}
where the definition domain of $\lambda$ includes the shoulder
bandpass(es). For a spectrum with {\it SNR} $> 20$, the typical
uncertainty in the EW of an atomic line is less than 0.1 \AA
\citep{2015RAA....15.1137L}. Note that for emission lines the relevant
EWs are negative, so the smallest value indicates the strongest line
strength.

The measured emission-line indices were accepted if $SNR_{\rm local}$
was greater than 5, and the number of bad pixels in the index bandpass
was zero. The resulting measurements are shown in Figure
\ref{fig:elements}.

The top two panels show the EW as a function of $\phi$ for H$\delta$
and H$\gamma$. The general trend of `higher luminosity, stronger
emission' for O-rich Miras is consistent with many previous case
studies \citep[e.g.,][]{1984ApJ...286..337F, 1985ApJ...297..455F,
  1995ASPC...83..419D, 2000AJ....120.2627C}. The bottom two panels
clearly show that the C-rich Miras exhibit stronger emission in
lower-order lines and particularly much greater variation in
H$\beta$. \citet{2005CoSka..35...83M} showed that for C-rich Miras,
the H$\alpha$ EW is also strongly correlated with photometric
phase. This can also be seen in the bottom right-hand panel of Figure
\ref{fig:elements}.

Our limited sample is based on observations of different stars over
different pulsation cycles, and it is thus not accurate enough to
derive strict relations. However, it can provide an approximate test
of how LAMOST spectra can support previous studies and provide us with
information regarding which emission lines should be measured for the
identification of O- and C-rich Mira candidates. Moreover, LAMOST has
been scheduled to observe more KISOGP Miras, whose spectra will allow
us to better constrain phases of light minima (or maxima), with higher
precision. These data will allow us to provide a more secure treatment
of the interaction between the propagation of shock waves and the
stars' pulsations.

\section{Candidate Selection}
\label{sec:identification}

In this section, we select possible Mira stars from the LAMOST DR4
catalog. We are motivated to do so with the expectation that such an
inquiry may lead to the discovery of a large number of Mira
candidates. The key steps are pre-selection based on photometric data,
followed by spectral-line measurements.

\subsection{Pre-selection}
\subsubsection{SNR Cut}

Based on the notion that Miras have stronger fluxes at red
wavelengths, we found the $i$-band {\it SNR} $> 20$ to be a good
diagnostic of high-quality Mira spectra. All of our 291 templates meet
this criterion.

\subsubsection{Header Subclass}

As outlined in Section \ref{subsec:lamost}, every LAMOST spectrum has
a spectral type assigned to it by the LAMOST 1D pipeline
\citep{2015RAA....15.1095L}. This pipeline produces four primary
classes, i.e., star, galaxy, qso, and unknown. Within the `star'
class, the `subclass' keyword provides a more detailed subtype. For
galaxies, quasars, or unknown objects, this keyword is set to `Non.'
To test the accuracy of the pipeline for Mira stars, we found that all
of our sample objects were included in the `star' or `unknown'
classes; the associated subclasses are included in Table
\ref{tab:subclass}.

\begin{deluxetable}{c|cccc}[ht!]
\tablecaption{Subclasses of our Mira sample identified by the LAMOST
  1D pipeline\label{tab:subclass}}
\tablehead{
\colhead{Type}& \colhead{Total}& \colhead{M}& \colhead{Non\tablenotemark{a}}&
  \colhead{Carbon}}
\startdata
Oxygen-rich & 258 & 219 & 38 &  1\tablenotemark{b}  \\
Carbon-rich & 33   & 1\tablenotemark{c}     &  5 & 27
\enddata
\tablenotetext{a}{Failure to identify a subclass.}
\tablenotetext{b}{This is a type-M9 spectrum, misclassified as a
  result of the small number of late-type M-giant templates.}
\tablenotetext{c}{This is an SC-type star, misclassified as a result of
  the absence of S-star templates.}
\tablenotetext{~}{~}
\end{deluxetable}

Given that 257 of the 258 (99.6\%) O-rich Mira templates are
identified as `M' or `Non,' we selected the 311,867 spectra of these
two subclasses as our O-rich Mira candidates. Although the automated
pipeline is not sufficiently efficient in classifying carbon stars,
\citet{2016ApJS..226....1J} already identified 108 N-type carbon stars
in the LAMOST DR2 catalog, and Y. B. Li et al. (in prep.) identified
266 N-type carbon stars in the DR4 catalog using different
techniques.\footnote{57 of the 266 carbon stars come from the LAMOST
  pilot survey; they are not included in the DR4 catalog.} We included
the results of these two studies and retrieved a total of 328 carbon
stars, which were then combined with the 2184 spectra marked as
`Carbon Star' by the LAMOST pipeline. We subsequently removed all
known Mira spectra from this database, yielding a total candidate
number of 313,810.

\subsubsection{2MASS Cuts}
\begin{figure} 
\center
\includegraphics [scale=0.42] {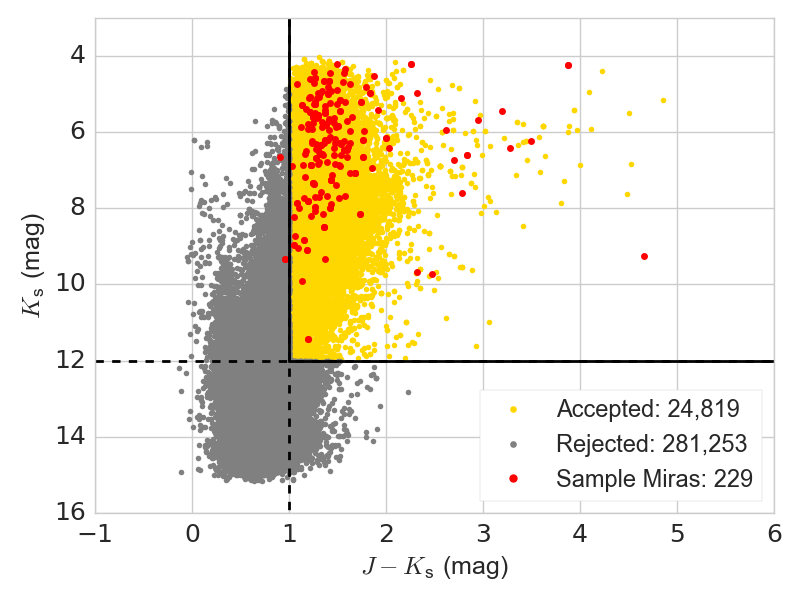}
\caption{2MASS color--magnitude diagram in the
  $\textit{K}\textsubscript{s}$ vs. ($J-$\textit{K}\textsubscript{s})
  plane. The red dots represent template Mira stars, while yellow and
  gray dots represent accepted and rejected candidates,
  respectively. The solid black lines of
  $J-$\textit{K}\textsubscript{s} $> 1$ mag and
  $\textit{K}\textsubscript{s} < 12$ mag illustrate the criteria
  adopted to select Mira candidates.\\
\label{fig:crude}}
\end{figure} 

Next, these Mira candidates were mapped onto the
\textit{K}\textsubscript{s} versus $J-$\textit{K}\textsubscript{s}
plane. $J$ and \textit{K}\textsubscript{s} magnitudes were obtained
through cross-identification with the Two Micron All-Sky Survey
(2MASS) all-sky point source catalog \citep{2006AJ....131.1163S}. Only
those stars with good photometric data (an `A' quality flag in the $J$
and \textit{K}\textsubscript{s} bands, i.e., 306,072 spectra) are
shown in Figure \ref{fig:crude}. Among the 291 Mira templates, 229
Miras with high-quality 2MASS photometry are also included in the same
plot. In Figure \ref{fig:crude}, 99.1\% of the templates are
sufficiently cool to be characterized by the empirical colors of giant
stars, $J-$\textit{K}\textsubscript{s} $> 1$ mag
\citep{2011A&A...530A...8W}, as well as bright,
$\textit{K}\textsubscript{s} < 12$ mag. Two outliers with
$J-$\textit{K}\textsubscript{s} $< 1$ are relatively bluer O-rich
Miras. Therefore, by applying careful cuts in the color--magnitude
diagram, we can remove most of the possible contaminators, leaving
32,557 (24,819 accepted plus 7738 without good 2MASS data) candidates.

\subsubsection{SIMBAD Data}

Finally, these candidates were cross-matched with the SIMBAD database;
1064 objects have SIMBAD designations that rule out a possible Mira
nature, thus leaving us with 31,493 candidates. In this step, we also
collected 197 spectra of known young stellar objects (YSOs) and T
Tauri stars (TTS), both of which are types of pre-main-sequence dwarf
stars. Their spectra will also be used in the next step.

\subsection{Selection of M Giants} \label{subsec:giants}

Contamination among the 30,248 O-rich candidates occurs mainly in the
form of M dwarfs and `Non'-marked spectra of other
types. \citet{2012ApJ...753...90M} showed that molecular band indices
of CaH2, CaH3, and TiO5 are good gravity indicators for the
determination of giant and dwarf luminosity classes. Introduced by
\citet{1995AJ....110.1838R}, band indices are calculated using the
ratio of the average flux over the specified wavelength range, i.e.,
 \begin{equation}
    R_{\rm ind} = \frac{\langle f_{\rm index} \rangle}{ \langle f_{\rm shoulder}\rangle}. 
    \label{eq:Rind}
 \end{equation}
Here, $\langle f_{\rm index} \rangle$ and $\langle f_{\rm
  shoulder}\rangle$ are the mean fluxes of all pixels in the index
bandpass and the shoulder bandpass defined in Table
\ref{tab:definition}, respectively. $R_{\rm ind}$ is a unitless number
between 0 and 1. It decreases as the absorption becomes stronger.

\begin{figure} 
\center
\includegraphics [scale=0.48] {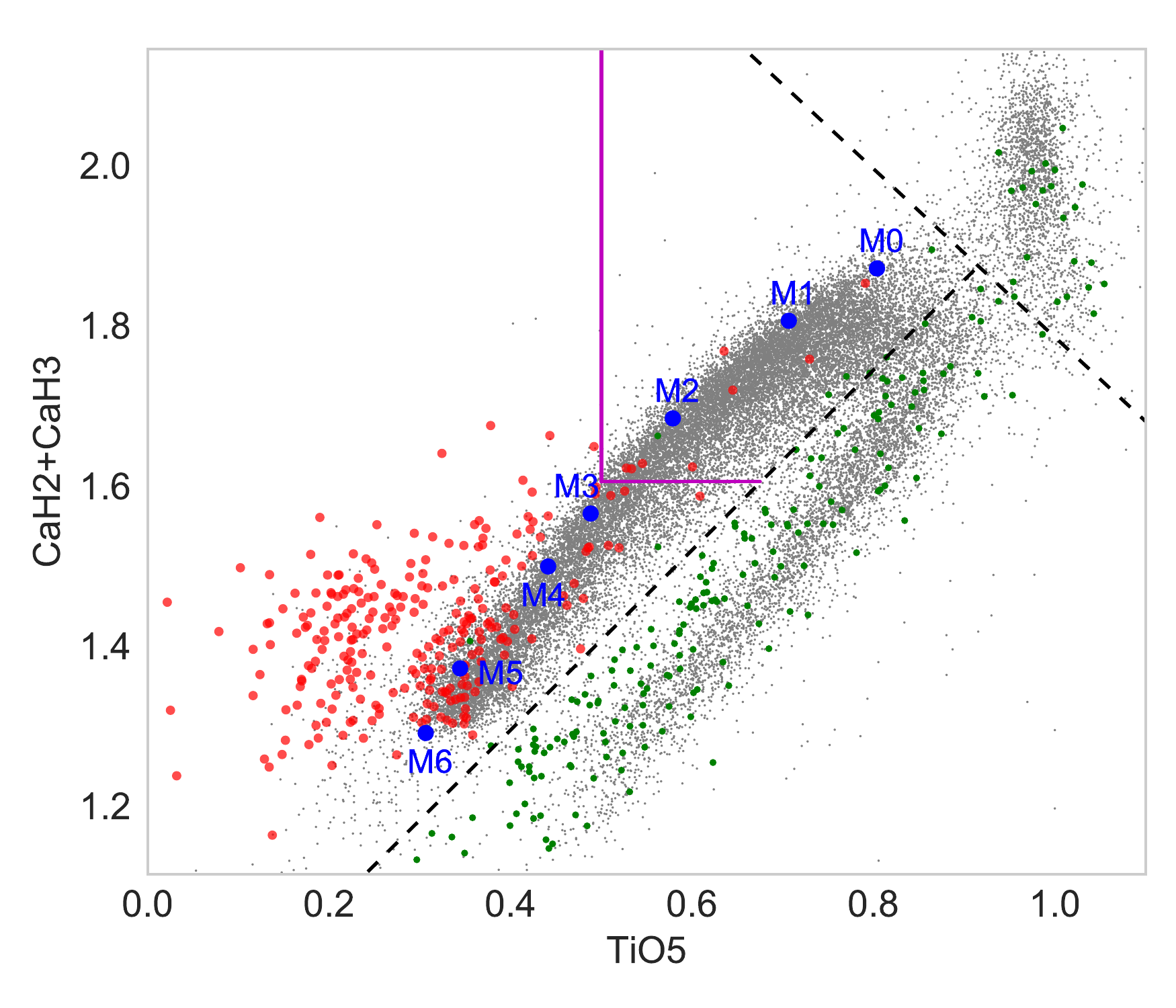}
\caption{Distribution of candidates (gray), Miras (red), M-giant
  templates from Z15 (blue), YSO and TTSs (green) spectra in the
  CaH2+CaH3 versus TiO5 diagram. The black dashed lines represent Eqs
  (\ref{eq:giant1}) and (\ref{eq:giant2}), which are used to select
  21,989 O-rich candidates and reject 8558 other candidates. The
  magenta lines represent Eqs (\ref{eq:magenta1}) and
  (\ref{eq:magenta2}).\\
\label{fig:two}}
\end{figure}

We map each candidate spectrum associated with a real
value\footnote{191 of the measured indices are `infinity' or `not a
  real number,' which are the results of bad pixels or corrupted
  spectra. We inspected these spectra by eye and rejected them as Mira
  spectra.} for the line index onto the CaH2+CaH3 versus TiO5 plane in
Figure \ref{fig:two}. Also included are the 258 O-rich templates (red)
and the M0--M6 normal M-giant templates (blue) of Z15. Band indices of
early-type giant stars are larger than those of late-type giants.

\citet{2011A&A...534A..79I} showed that some TTSs and YSOs have colors
similar to O- or C-rich AGB stars. In addition, the optical spectra of
M-type YSOs and TTSs \citep{2003A&A...409..993S, 2014ApJ...786...97H}
also have strong Balmer emission lines. To test if the CaH2+CaH3
vs. TiO5 diagram can efficiently separate them from M giants, 197
spectra of YSO and TTSs (green) are shown in green.

Compared with M dwarfs, M giants are located in the top branch because
of their weaker CaH molecular bands, and they are clearly
distinguishable by a selection cut of
\begin{equation}
{\rm CaH2}+{\rm CaH3} > 1.125 \times {\rm TiO5} + 0.85. \label{eq:giant1}
\end{equation}
In the top right-hand region of Figure \ref{fig:two}, a cluster of
dots centered on (TiO5 = 1, CaH2+CaH3 = 2) do not have any CaH or TiO
features and thus are not spectra of M-type stars. A small number of
YSO and TTSs also reside in this region; these are F- or G-type
stars. They should be excluded from our O-rich candidates by a
selection cut of
\begin{equation}
 {\rm CaH2}+{\rm CaH3} < -{\rm TiO5} + 2.8. \label{eq:giant2}
\end{equation}
After having applied the cuts of Eqs (\ref{eq:giant1}) and
(\ref{eq:giant2}), the number of O-rich candidates has been reduced to
21,989. Most TTSs and YSOs are excluded by application of the two
cuts.

Following the distribution of M0--M6 templates, we roughly separated
the accepted 21,989 candidates into 12,719 early- and 9270 late-type
giants using cuts of
\begin{equation}
{\rm CaH2+ CaH3 }> 1.61 \label{eq:magenta1}
\end{equation}
and
\begin{equation}
{\rm TiO5} > 0.50.\label{eq:magenta2}
\end{equation}
The intention of this separation is to leave spectra of M5--M10 for
the next step, i.e., the measurement of metallic lines, because based
on Section \ref{subsubsec:metallic emission}, we do not expect such
emission to be present in warm O-rich spectra. We intentionally
adopted a flexible and wide limit (retaining M3--M10 instead of
M5--M10), since we allow for contamination of early-type spectra in
the late-type group.


\subsection{Fe and Mg Emission} \label{subsec:Fe/Mg emission}
\begin{figure} 
\center
\includegraphics [scale=0.42] {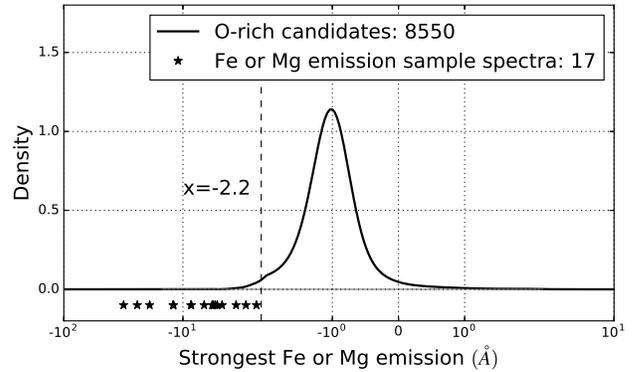}
\caption{Distribution of The Strongest Emission EWs in Fe{\sc i}
  $\lambda\lambda$4202,4308,4376 and Mg{\sc i} $\lambda$4571. Kernel
  Density Estimation (KDE) with a Gaussian kernel is used for the
  visualization. $X$-axis values of the black asterisks at the bottom
  are EWs of all Fe and Mg emission template spectra.
\label{fig:Fe}}
\end{figure}

As suggested by \citet[][their Figure 23]{2001A&A...369.1027R} and
\citet[][page 304]{2009ssc..book.....G}, in O-rich Mira stars the
metallic emission strength is not in phase with the Balmer strength.
Therefore, before we measure the most commonly observed Balmer
emission lines, line indices of Fe{\sc i}
$\lambda\lambda$4202,4308,4376 and Mg{\sc i} $\lambda$4571 were
measured to identify Mira spectra with Fe or Mg emission, which might
not display Balmer lines.

Our candidates are the 9270 late-type O-rich spectra obtained in the
previous step, and our templates are the 20 emission spectra shown in
Figure \ref{fig:femg}. Using Eq. (\ref{eq:EW}) as our definition of
the EW and Eq. (\ref{eq:SNR}) to quantify the local SNR, we only
accepted EWs if their $SNR_{\rm local}$ was no less than five, and if
there was no bad pixel in the index bandpass, which means that we used
the same criteria as those adopted in Section
\ref{subsec:sample_emission}.

Figure \ref{fig:Fe} shows the probability density distribution of the
strongest measured Fe or Mg EWs of our late-type, O-rich
candidates. Here we choose the smallest value among the four measured
line indices to single out the strongest emission. This might
potentially bias the line indices to negative values for a given
spectrum containing other spectral features or only noise across the
measured wavelength range. Another potentially biasing effect is
associated with an absorption band at 4582 \AA, near the right-hand
shoulder of Mg{\sc i} $\lambda$4571, which might cause an artificial
emission feature and systematically drive the measured value to become
smaller. Therefore, the peak of the profile in Figure \ref{fig:Fe} is
centered on minus one rather than zero.

Since all of the 20 Fe or Mg emission spectra\footnote{17 have
  `accepted' strongest EWs.} among our templates have ${\rm EW}({\rm
  Strongest \ Fe \ or \ Mg})<-2.4$ \AA, we adopted a cut at
\begin{equation}
{\rm EW(Strongest \ Fe \ or \ Mg)} < -2.2 {\rm \AA} \label{eq:Fe}
\end{equation}
to select a sample composed of 338 candidates. We inspected the 338
spectra by eye. Most of their negative line indices are caused by
noise. However, we also found 14 spectra in the range of M7--M10 that
should be classified as Mira stars, as shown in the Appendix (see
Figure \ref{fig:metal_selected}). Three of these 14 spectra have
prominent Fe or Mg lines (two with Balmer lines and one without), and
the other 11 are classified based on the presence of hydrogen
emission.

Now that we have successfully selected 14 Mira spectra, the number of
O-rich candidates is reduced to 21,651.


\subsection{Balmer Emission}\label{subsec:Balmer_selection}
\begin{deluxetable*}{ccccc}[htp]
\tablecaption{Steps to Identify Mira stars \label{tab:steps}}
\tablehead{
\colhead{Step}& \colhead{\ \ \ \ \ Criteria \ \ \ \ \ }& \colhead{\ O-rich\ } &\colhead{C-rich} &\colhead{Total \# candidates}
} 
\startdata
     1                                  & $SNR(i)>20$                                                                     &                       &                &5,025,780\\
     2			                & LAMOST Pipeline subclass				                 & 311,867         & 2184       &314,051   \\
     3                                  & + Previously identified N-type star                    		& 311,867      	& 2314       & 314,181  \\                                      
     4                                  & -- Known Mira spectra							& 311,537      	& 2273       & 313,810   \\
     3                         & $J-K_{\rm s}$ \textgreater1 mag, $K_{\rm s}$\textless12 mag  & 31,795        	& 762         & 32,557	 \\
     4                                  & -- SIMBAD type: not a Mira 	                         			& 30,738        	& 755         & 31,493	\\
     5                                  & Eqs (\ref{eq:giant1}), (\ref{eq:giant2}) 				& 21,989        	& 755         & 22,744	\\        
     6                                  & Eq. (\ref{eq:Fe})                                  				& 21,651(+14)	& 755         & 20,542\\ 
     7                                  & Eq. (\ref{eq:H})                                                            	& 937(+14)  	& 263         &1214 \\
     8                                  & Inspection by eye                                                           	& 106(+14)	& 93           & 213 \\
     9                                  & Single epoch                                                                     &  110		& 81           & 191
\enddata
\end{deluxetable*}

\begin{deluxetable*}{cccccccccccc}
\tablecaption{Mira Stars Discovered in LAMOST DR4 \label{tab:selected}} 
\tablecolumns{12} \tablewidth{0pt}
\tablehead{ \colhead{R.A. (J2000)} & \colhead{Dec. (J2000)} & \colhead{Designation} &
  \colhead{Epoch\tablenotemark{a}}& \colhead{SIMBAD\tablenotemark{b}}
  & \colhead{$J$} & \colhead{$H$} & \colhead{$K$\textsubscript{s}} &
  \colhead{$\sigma_{J}$\tablenotemark{c}} & \colhead{$\sigma_{H}$} &
  \colhead{$\sigma_{K_{\rm s}}$} &
  \colhead{SpType\tablenotemark{d}}\\ \colhead{degree} &
  \colhead{degree} & \colhead{} & \colhead{} & \colhead{} &
  \colhead{mag} & \colhead{mag} & \colhead{mag} & \colhead{mag} &
  \colhead{mag} & \colhead{mag} & \colhead{} } 
\startdata 
30.03712   & 41.62985 & J020008.90+413747.4 & 56256 & C*                                         & 10.867 & 9.549 &8.656 & 0.024 & 0.003 & 0.02  & Carbon \\ 
301.53141 & 46.14646 & J200607.53+460847.2 & 56560 57298 \tablenotemark{e} & C* & 8.85    & 7.803 &7.327 & 0.029 & 0.026 & 0.023 & Carbon \\ 
78.49842   & 30.11585 & J051359.62+300657.0 & 55858 55909 & & 8.902 & 7.931 & 7.44 & 0.026 & 0.026 & 0.02 & M5, M4\tablenotemark{f} \\ 
15.59558   & 37.91861 & J010222.93+375506.9 & 55907 & LPV* & 7.432 & 6.546 & 6.16 & 0.03 & 0.029 & 0.02 & M7 \\ 
93.02157   & 12.11717 & J061205.17+120701.8 & 57388 & IR & 7.291 & 5.999 & 5.195 & 0.023 & 0.034 & 0.018 & M8 
\enddata 
\tablenotetext{a}{LAMOST Modified Julian Day (MJD).}
\tablenotetext{b}{SIMBAD main type.}
\tablenotetext{c}{2MASS photometric uncertainty.}
\tablenotetext{d}{Spectral subtype.}
\tablenotetext{e}{LAMOST observed this target twice.}
\tablenotetext{f}{Temperature type of different observations.}
\tablecomments{This table is available in its entirety in machine-readable form.}
\end{deluxetable*}

\begin{figure}
\includegraphics [scale=0.42] {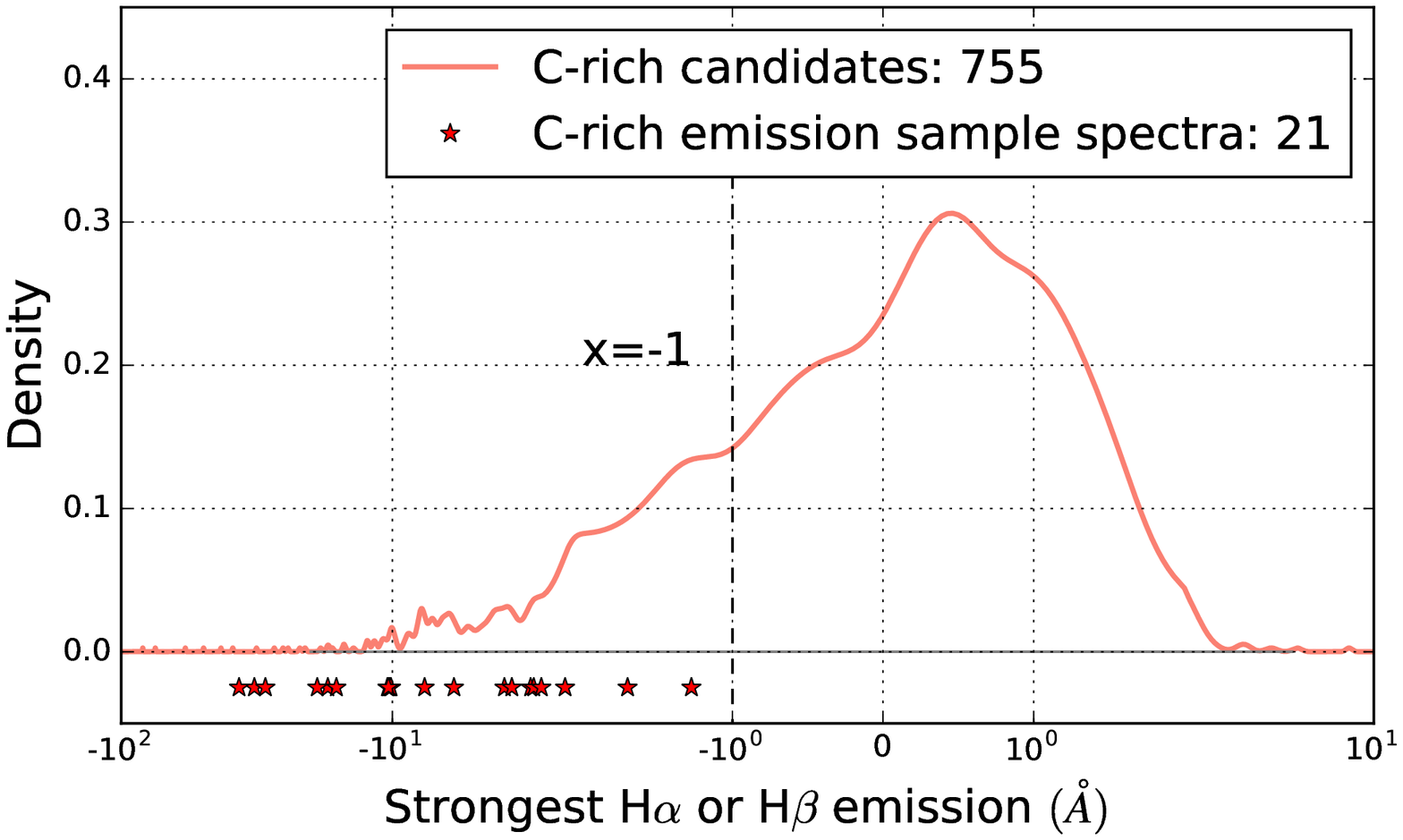}\\
\includegraphics [scale=0.42] {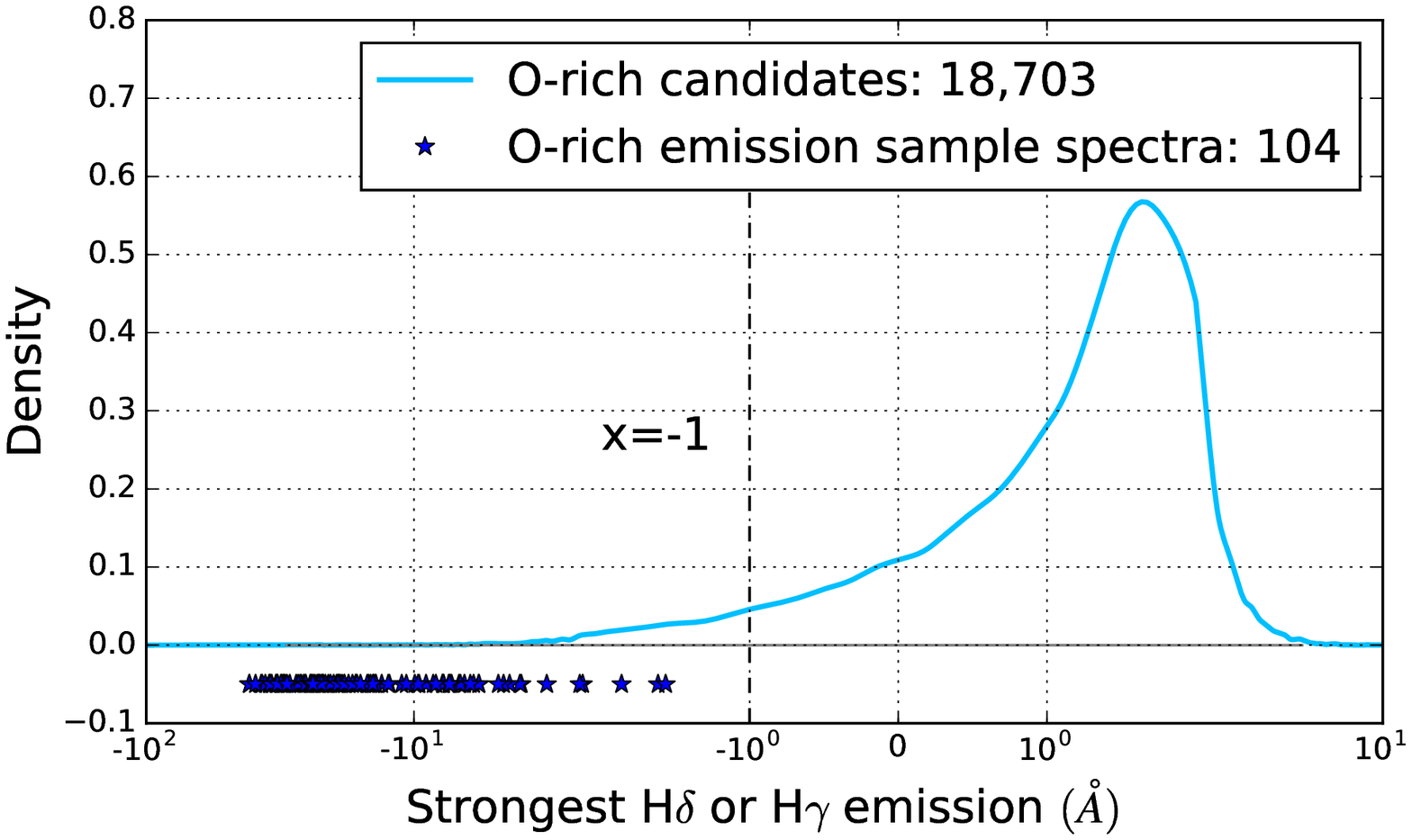}
\caption{Top (bottom): Distribution of The Strongest Emission EWs in
  H$\alpha$ and H$\beta$ (H$\gamma$ and H$\delta$). The probability
  density distribution of C-rich (O-rich) candidates is shown by the
  KDE with a Gaussian kernel of 0.2 (0.1) times the bandwidth in the
  top (bottom) panel.
\label{fig:emission}}
\end{figure} 

EWs of hydrogen emission lines need to be measured to identify Mira
spectra, see Eq. (\ref{eq:EW}), Table \ref{tab:definition}. We chose
to measure the first two lines in the Balmer series (H$\alpha$ and
H$\beta$) for our 755 C-rich candidates, and the next two lines
(H$\gamma$ and H$\delta$) for the 21,651 O-rich candidates. Again, we
adopted the smallest value as an indication of the emission
features. The acceptance criteria for every measurement are also the
same as those adopted in Sections \ref{subsec:sample_emission} and
\ref{subsec:Fe/Mg emission}. However, this time we do not retain the
`rejected' spectra in our next step, because a high {\it SNR} value in
the Balmer region is essential for identification of Mira
spectra. Balmer emission template spectra (21 C-rich and 104 O-rich)
were measured at the same time. A more plausible diagnostic might be
found in measuring H$\alpha$ and H$\delta$ for C- and O-rich
candidates, respectively. However, we measure two lines for each type,
because if the $SNR_{\rm local}$ of one line is low, we could
potentially still use the other. In addition, one C-rich template
spectrum only exhibits H$\beta$ emission (see Figure \ref{fig:cs2}).

In each panel of Figure \ref{fig:emission}, the asterisk closest to
zero is measured from a template spectrum whose Balmer emission can
just be seen, and this can thus give us an upper limit to a
distinguishable emission line. The peak of each profile is centered on
a positive value, because for normal M giants and carbon stars, the
Balmer lines are seen in absorption. For each panel a cut at
\begin{equation} 
{\rm EW(Strongest \ Balmer)} < -1 {\rm \AA}, \label{eq:H} 
\end{equation} 
was adopted. We were thus left with 937 O-rich and 263 C-rich
candidates, which were subsequently inspected by eye.


\subsection{By-Eye Inspection}\label{subsec:by_eye}

By-eye inspection can help us to decide whether a negative EW value
represents real emission or just spectral noise. The major difficulty
resides in identifying C-rich Mira spectra, because for normal C-rich
stars of N spectral type, H$\alpha$ tends to be filled in or weak in
emission. We rejected spectra with deep Ca{\sc ii}
$\lambda\lambda$8498,8542,8662 absorption features (Section
\ref{study_c}). In addition, we tried to make sure that all selected
C-rich spectra exhibited strong H$\alpha$ or H$\beta$ emission. The
199 spectra identified in this step are shown in the Appendix (see
Figures \ref{fig:o_selected} and \ref{fig:c_selected}).

We have summarized the full selection process in Table
\ref{tab:steps}. Our final sample of 191 identified Mira variables is
provided in Table \ref{tab:selected}.

\subsection{Spatial Distribution}
\begin{figure}
\center
\includegraphics[scale=0.43]{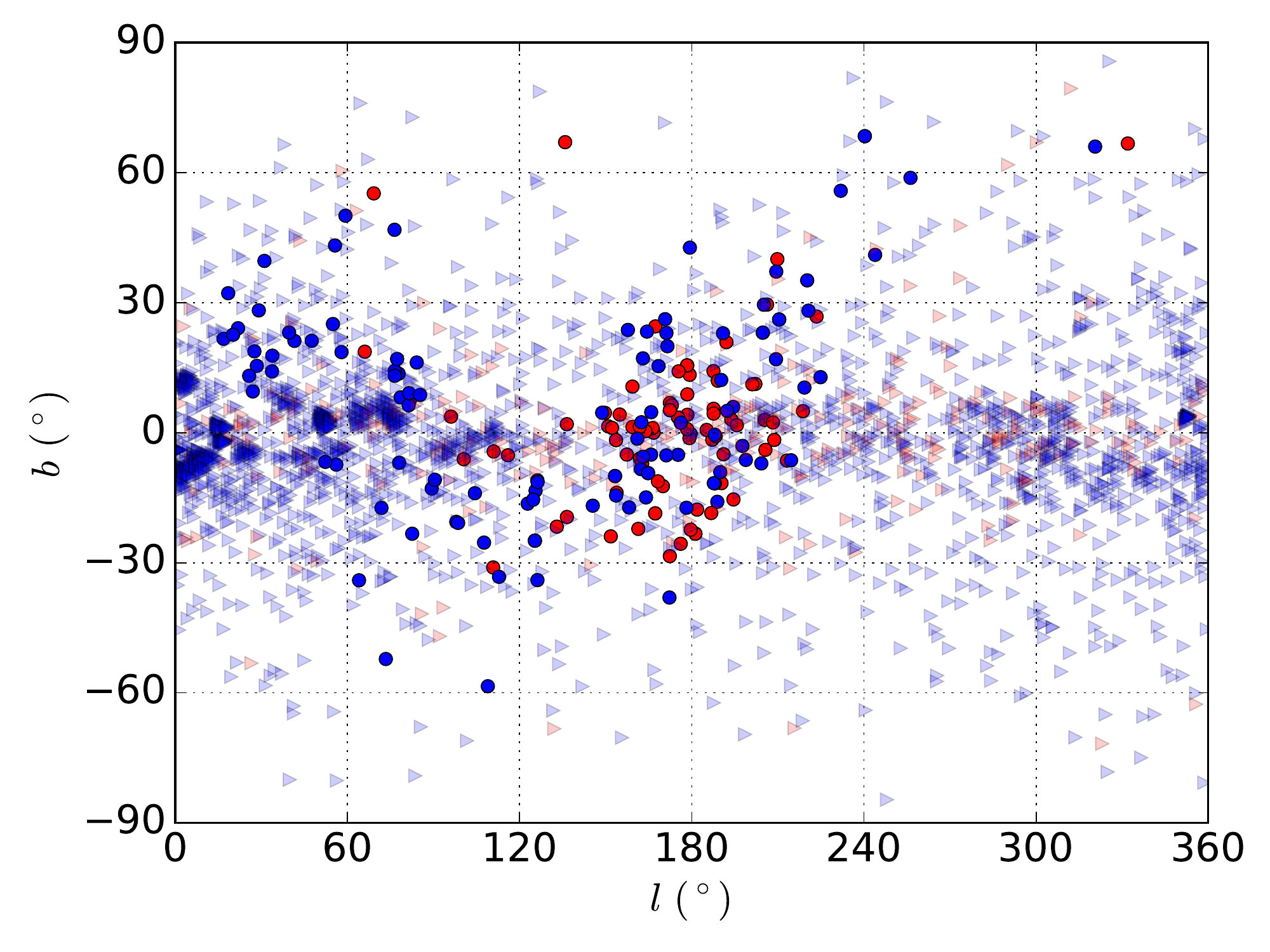}
\caption{Galactic distribution of Mira stars. M- and C-type Mira
  variables are plotted in blue and red, respectively. Transparent
  triangular data points are Miras in the AAVSO VSX for which a
  designated SIMBAD type is available; circles are Miras identified in
  this paper. The concentration of a group of O-rich Miras around $l
  \sim 75\degr$ is centered on the {\sl Kepler} field.
\label{fig:spatial}}
\end{figure}

Figure \ref{fig:spatial} shows the spatial distribution of the newly
identified Mira candidates in Galactic coordinates. Among the 2724
AAVSO Miras with available SIMBAD subtypes, the C- to O-rich ratio is
only 0.13.\footnote{We did not include S stars, because they can be
  either O- or C-rich, and the total fraction of S-type Miras is
  small.} As regards the 191 Miras identified using LAMOST, this ratio
increases to 0.74. This difference might be the result of a difference
between the spatial distributions of the different types of AGB
stars. It has been reported that O-rich AGB stars are concentrated
toward the Galactic Center, whereas C-rich AGB stars exhibit a
relatively uniform distribution \citep{2011A&A...534A..79I}, including
a few C-rich Miras recently confirmed in the bulge
\citep{2017MNRAS.469.4949M}. Since the target selection of the LAMOST
survey is not spatially uniform, but has a concentration of stars in
the Galactic Anticenter direction \citep{2012RAA....12..735D,
  2014IAUS..298..310L}, our selected Miras contain larger numbers of
carbon stars, and they are thus distributed more commonly toward the
$l \sim 180\degr$ region.


\section{Discussion and Conclusion}
\label{sec:conclusion}

In this work, we have carefully studied 291 LAMOST spectra of known
Mira stars to explore the optical properties of Miras. For O-rich
stars, a relationship between the relative Balmer emission-line
strength and spectral temperature is clearly shown. The $F_{\rm
  H\delta}/F_{\rm H\gamma}$ emission-line flux ratio increases as
stars cool down from M0 to M10. Spectra earlier than M4 exhibit
stronger H$\gamma$ than H$\delta$, which is a new result reported
here. This relation can be explained by overlying TiO molecular-band
absorption affecting lower-order Balmer lines. Metallic emission of
Fe{\sc i} $\lambda\lambda$4202,4308,4376 and Mg{\sc i} $\lambda$4571
is only observed in late-type stars.

Based on the well-established relation between Balmer emission-line
strength and the corresponding pulsation phase, our result shows that
such a relation is indeed a population property exhibited by Mira
variables. Future LAMOST observations can place stronger constraints
on the rise duration and maximum phase of H$\delta$ and H$\gamma$ for
O-rich Miras.

Based on the characteristics of the 291 spectral templates, we
identified 191 Mira candidates in the LAMOST DR4 catalog. Our criteria
to select Mira stars make the most of a wide variety of available data
(AAVSO VSX, SIMBAD, KISOGP, 2MASS, pipeline subclasses, spectral
flux), and this is thus an efficient approach in terms of time
commitment. Meanwhile, we made sure that every rejection was carefully
assessed and that inaccurate measurements were not used. In terms of
completeness, if we apply our the method outlined in Section
\ref{sec:identification}, 135 of the 291 template spectra we studied
in Section \ref{sec:templates} can be selected. About half of the
templates are rejected based on spectroscopic (null emission at a
given phase) or photometric (bluer than the bulk of other Miras)
criteria. Therefore, we estimate that the DR4 catalog may include a
total of $191\times \frac{291}{135} \simeq 412$ new Miras, with some
$412-191=221$ remaining undetected.

Note that since Mira variables are defined by their luminosity
variability, photometric observations are required to further confirm
that the newly identified candidates are bona fide Mira stars. It is
worth noting that six candidates have good photometric data and are
located within the KISOGP field. The light curves of three candidates
look like Miras (and all of them show long-term trends); two are found
near the boundary between Miras and semi-regular variables, while the
other is probably not a Mira. From this analysis we conservatively
expect the ratio of bona fide Mira stars to the number of selected
candidates to exceed 50\%.

Newly discovered Mira candidates should usefully contribute to studies
of stellar evolution of late-type stars (especially to shock-induced
pulsation studies of C-rich stars) in the Galactic Anticenter. Eight
of the 111 O-rich candidates are of type M0--M2, which are rare
cases. This speaks volumes for LAMOST's ability to capture the spectra
of early-type Miras through such a `treasure hunt'. Also, at least 24
of the 191 Miras are located in the Galactic halo (Galactic latitude
$>$ 30\degr) and can possibly be used as tracers of stellar streams,
although their low space density may prevent us from investigating
small structures.

\acknowledgements 
We acknowledge with pleasure the anonymous reviewer who has offered
valuable suggestions. This work was supported by the Hui-Chun Chin and
Tsung-Dao Lee Chinese Undergraduate Research Endowment
(CURE). C.L. acknowledges support from both the National Key Basic
Research Program of China (grant 2014CB845700) and the National
Natured Science Foundation of China (NSFC; grants 11373032 and
11333003). R.d.G. acknowledges NSFC funding through grants U1631102,
11373010, and 11633005. The Guoshoujing Telescope (the Large Sky Area
Multi-Object Fiber Spectroscopic Telescope, LAMOST) is a National
Major Scientific Project constructed by the Chinese Academy of
Sciences. Funding for the project was provided by the National
Development and Reform Commission. LAMOST is operated and managed by
the National Astronomical Observatories, Chinese Academy of Sciences.


\appendix
\begin{figure*}
\center
\subfigure[]{%
\begin{tabular}{c}
\includegraphics [scale=0.67] {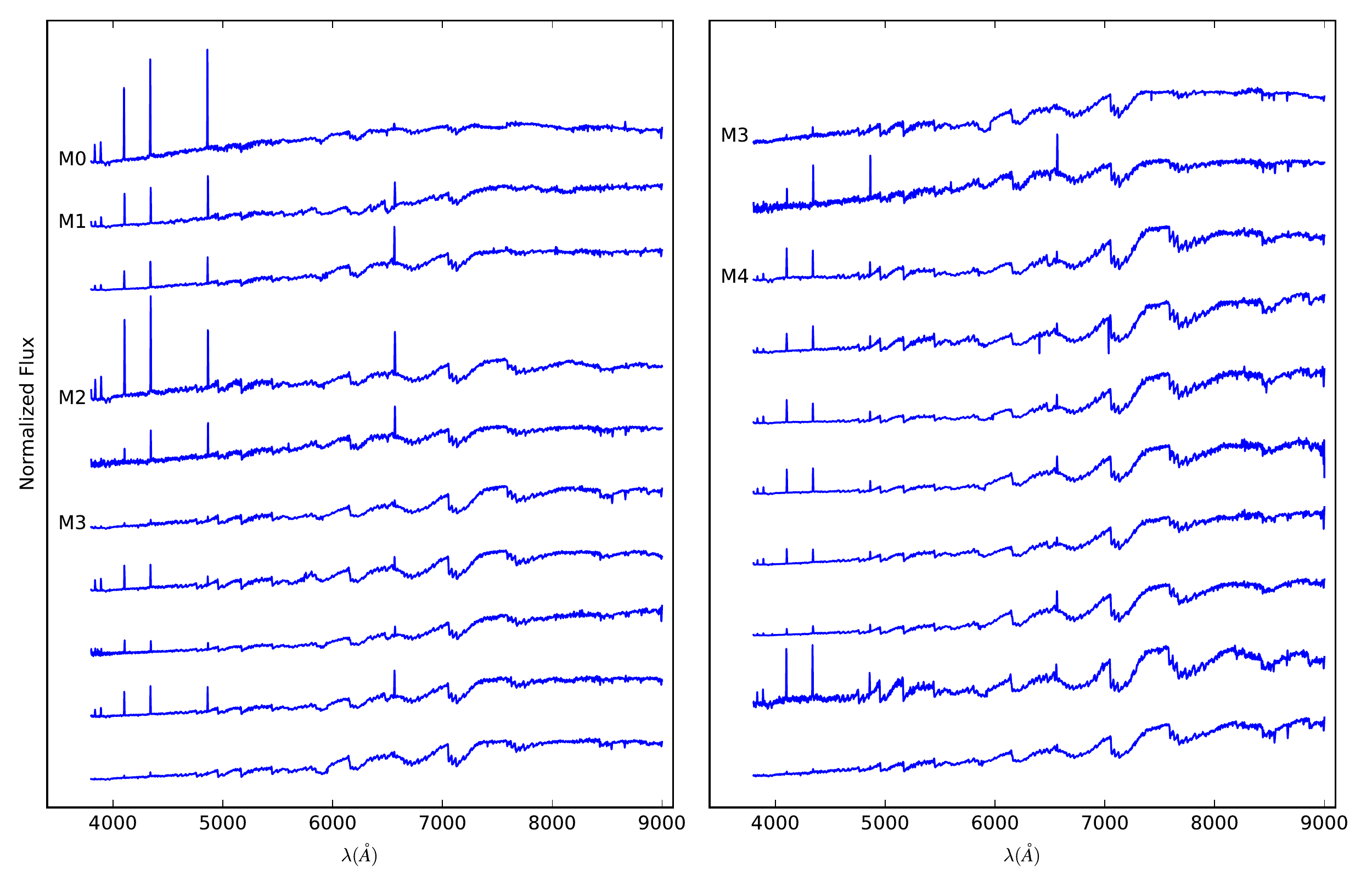}\\
\includegraphics [scale=0.67] {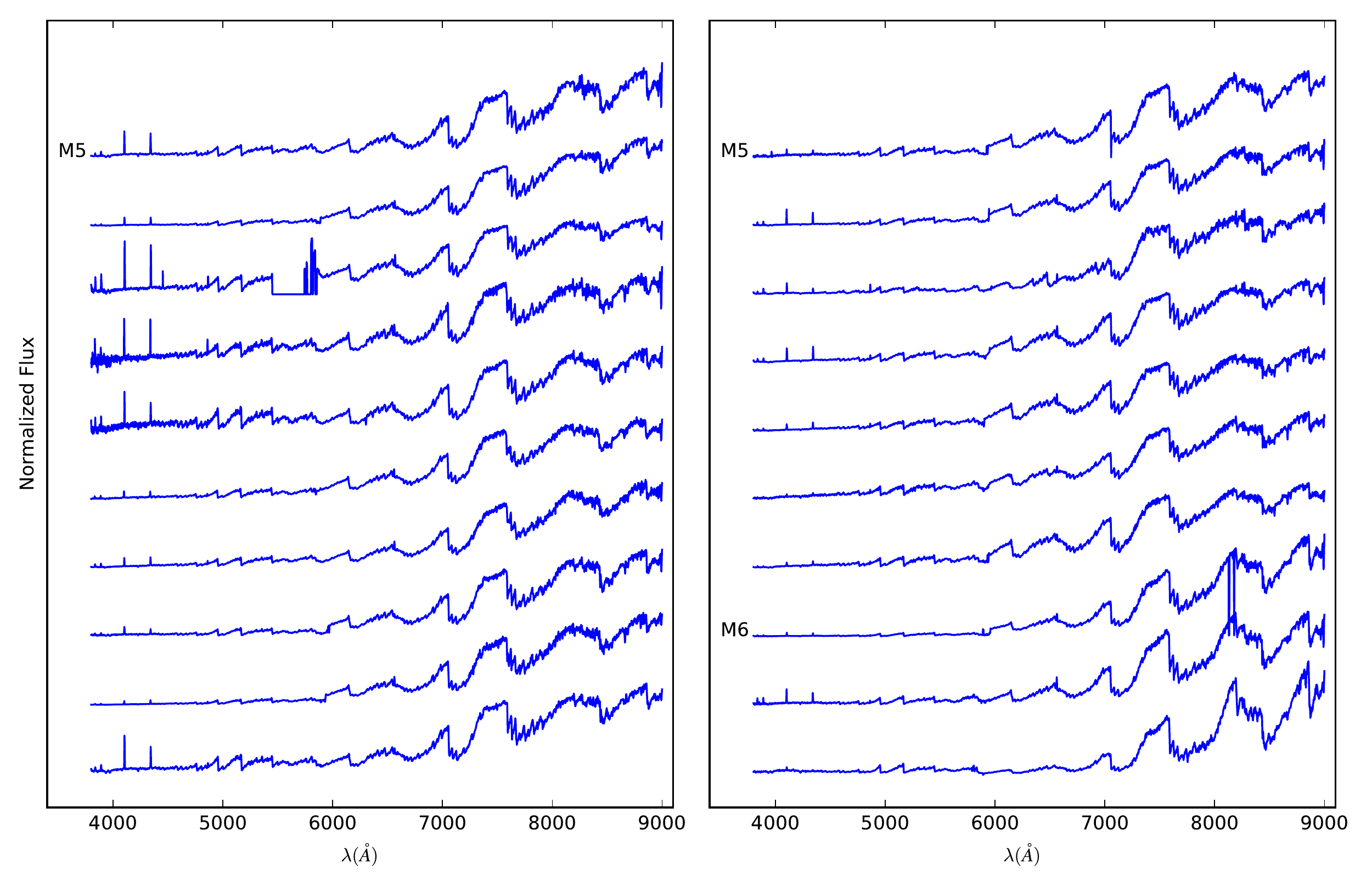}\\
\end{tabular}
}
\end{figure*}
\clearpage

\begin{figure*}
\center
\subfigure[]{%
\begin{tabular}{c}
\includegraphics [scale=0.67] {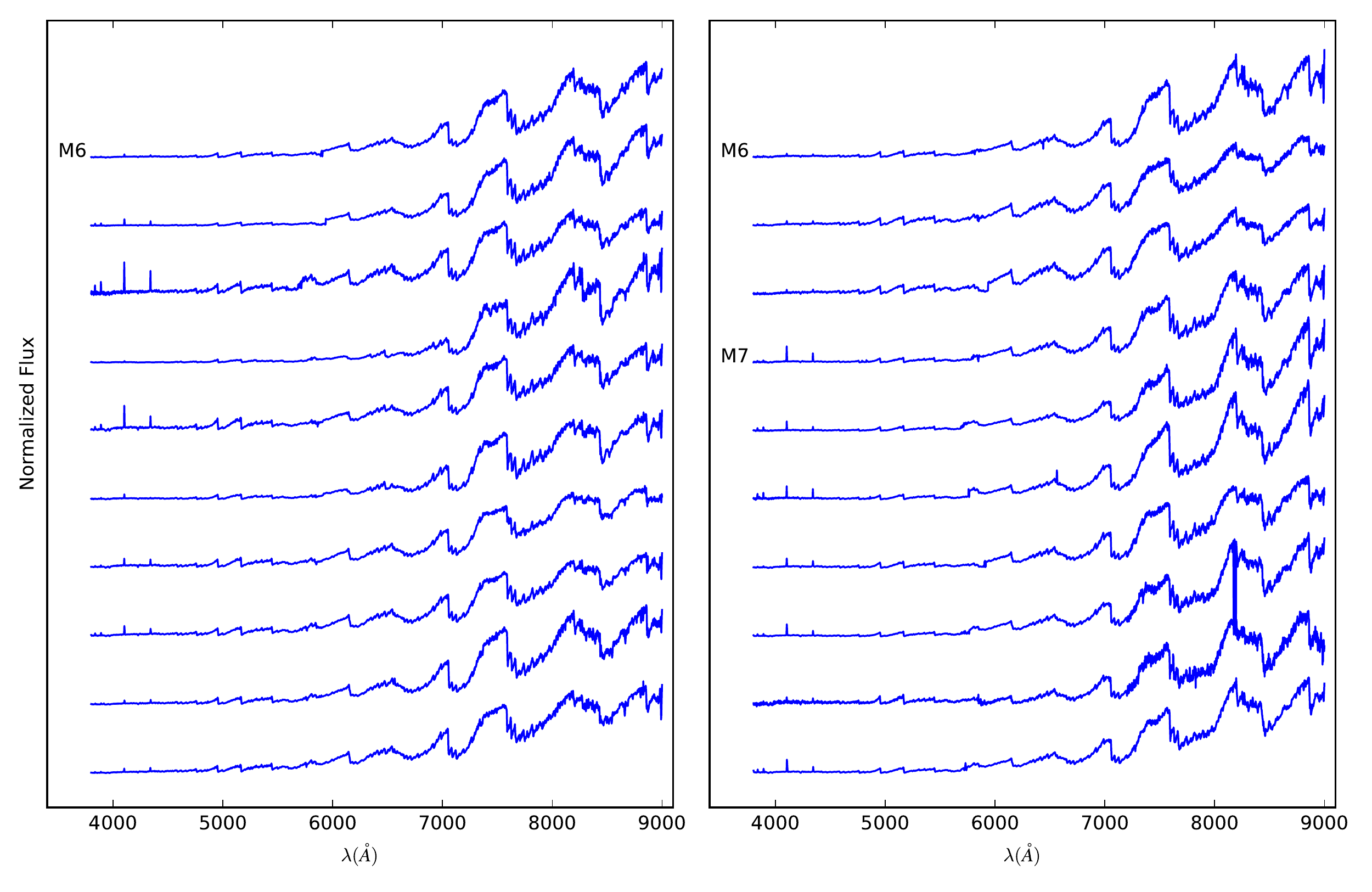}\\
\includegraphics [scale=0.67] {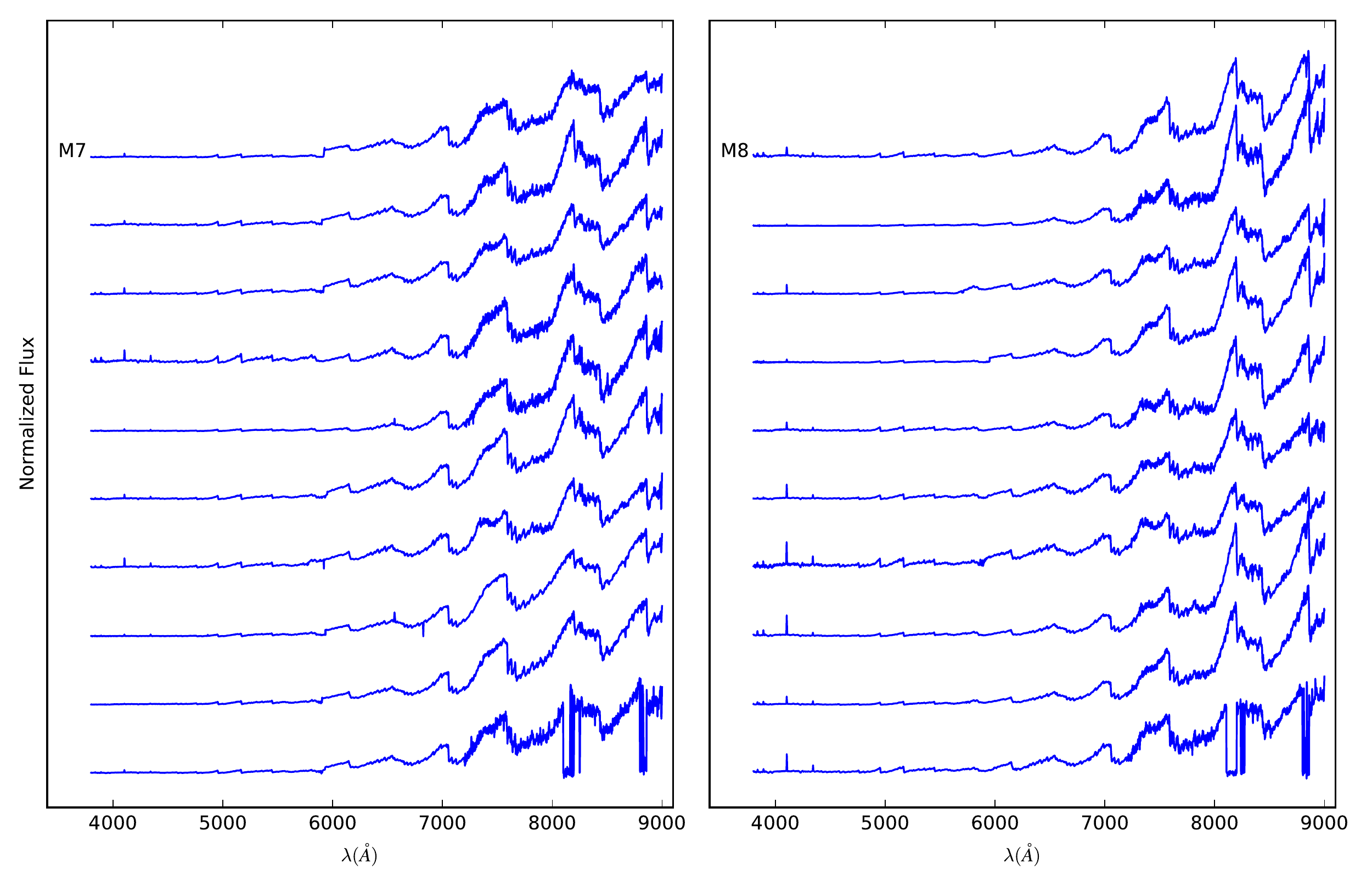}\\
\end{tabular}
}
\end{figure*}
\clearpage

\begin{figure}
\subfigure[]{%
\begin{tabular}{c}
\includegraphics [scale=0.67] {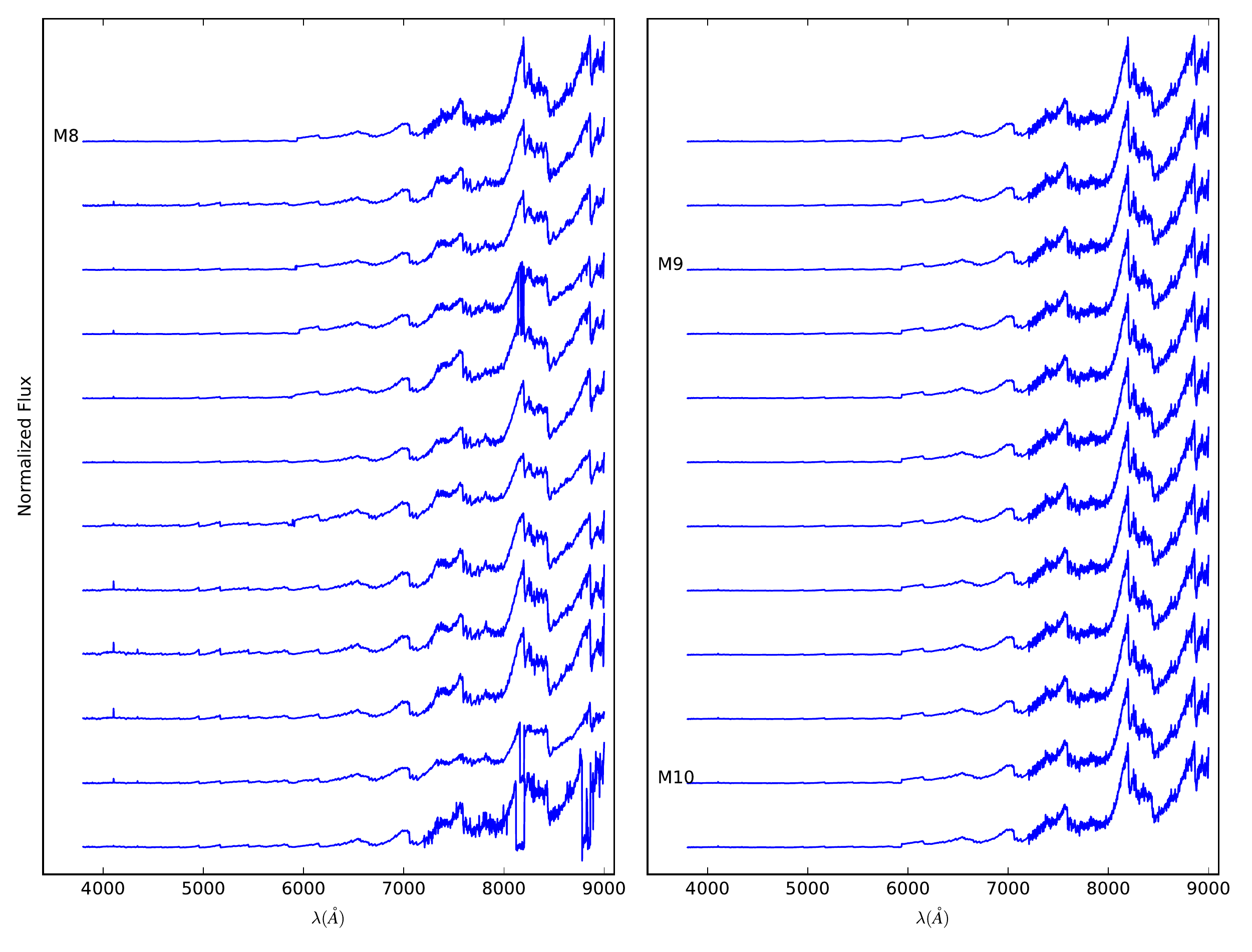}
\end{tabular}
}
\caption{Spectra of O-rich Mira templates, only uncorrupted spectra
  with emission lines are shown.
\label{fig:app5}}
\end{figure}

\begin{figure*}
\center
\subfigure[]{%
\begin{tabular}{c}
\includegraphics [scale=0.67] {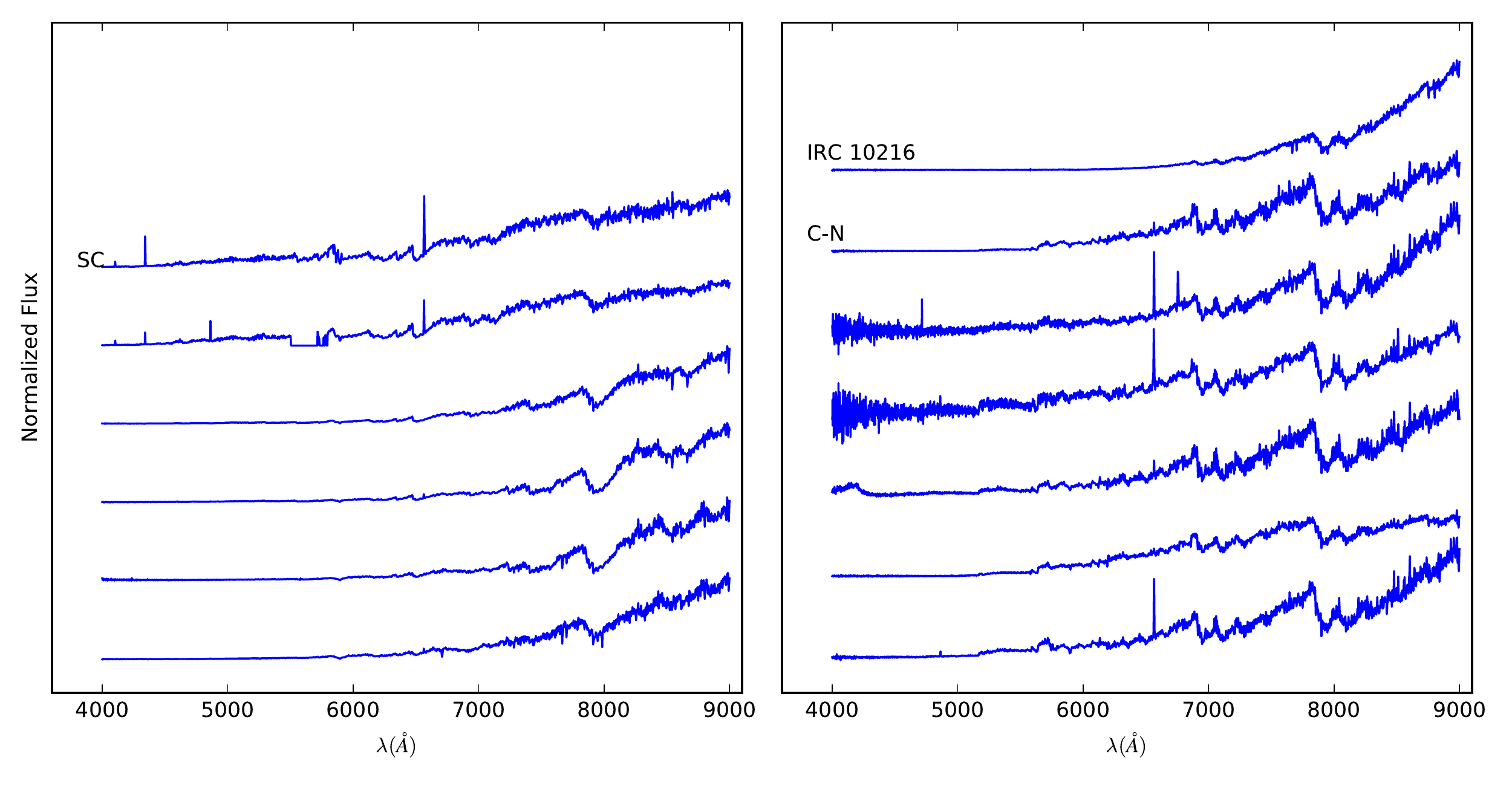}\\
\end{tabular}
}
\end{figure*}
\clearpage

\begin{figure*}
\center
\subfigure[]{%
\begin{tabular}{c}
\includegraphics [scale=0.67] {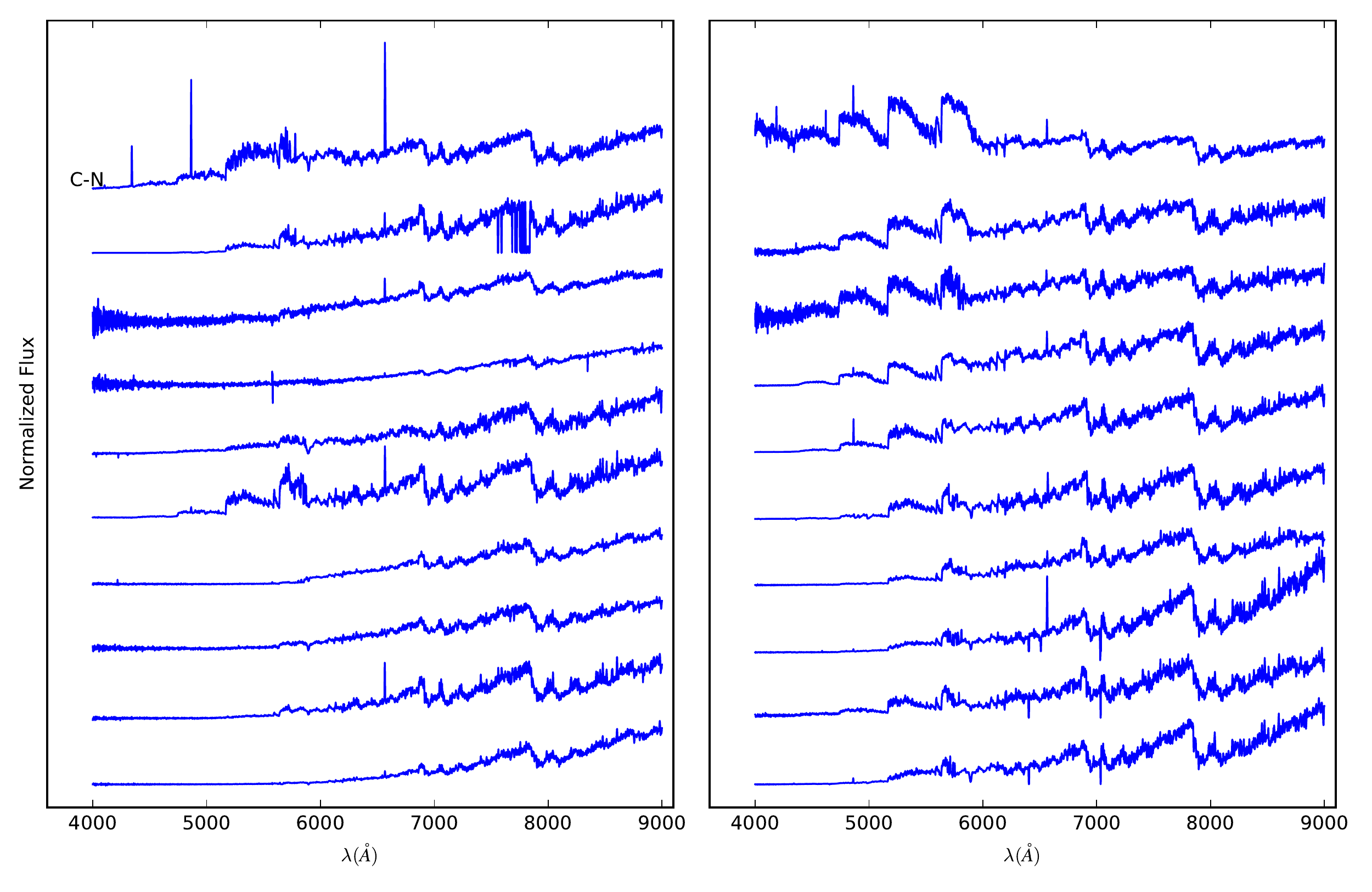}\\
\end{tabular}
}
\caption{Spectra of C-rich Mira candidates. The widely studied carbon
  star IRC 10216 is marked because of its redness.
	\label{fig:cs2}}
\end{figure*}
\begin{figure*}
\center
\begin{tabular}{c}
\includegraphics [scale=0.57] {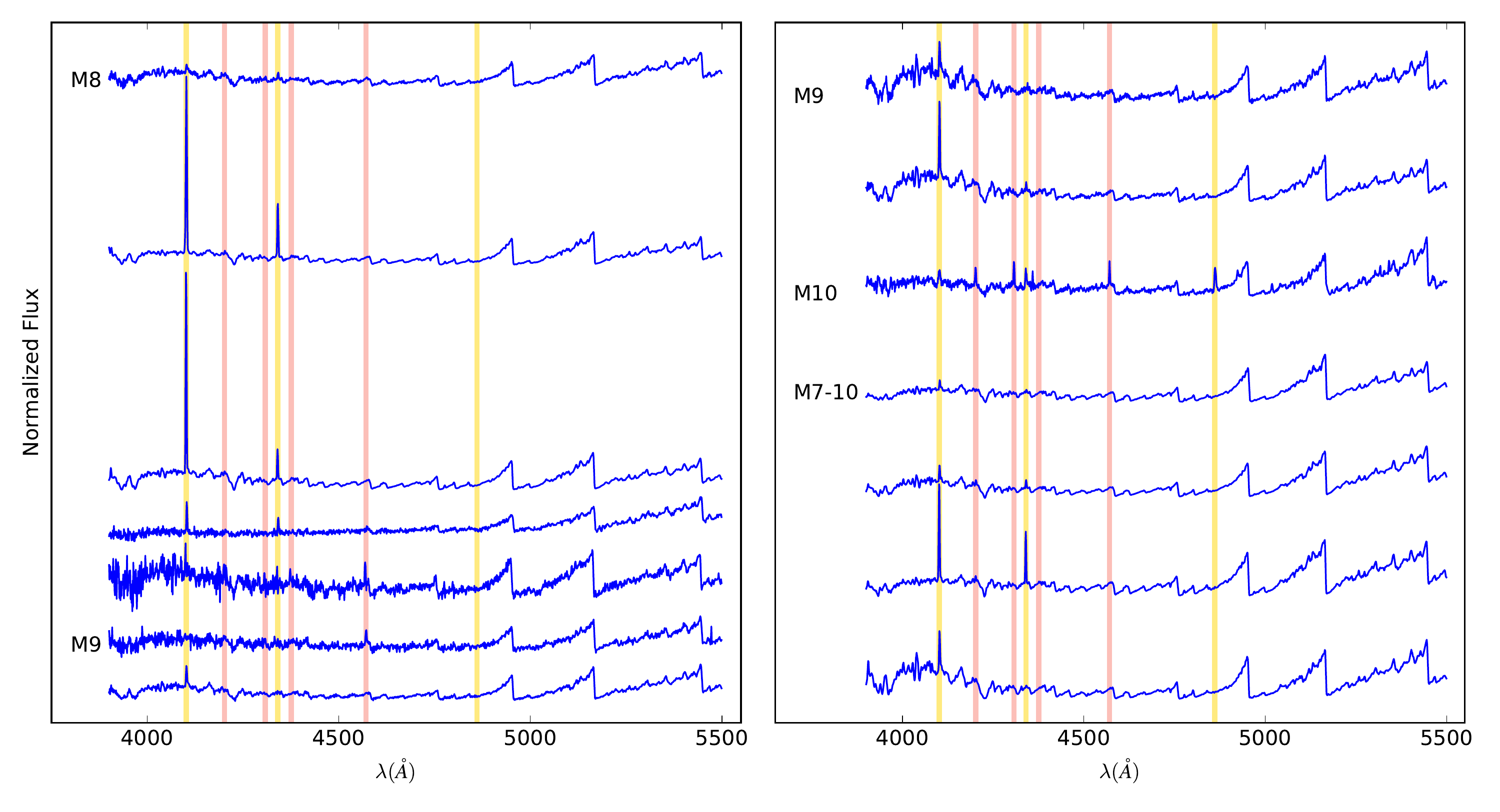}\\
\end{tabular}
\caption{Spectra selected in Section \ref{subsec:Fe/Mg emission}.
  Temperature types of foour spectra can only be roughly determined as
  M7--M10, because their red sections suffer from saturation.
\label{fig:metal_selected}}
\end{figure*}
\clearpage

\begin{figure*}
\center
\subfigure[]{%
\begin{tabular}{c}
\includegraphics [scale=0.67] {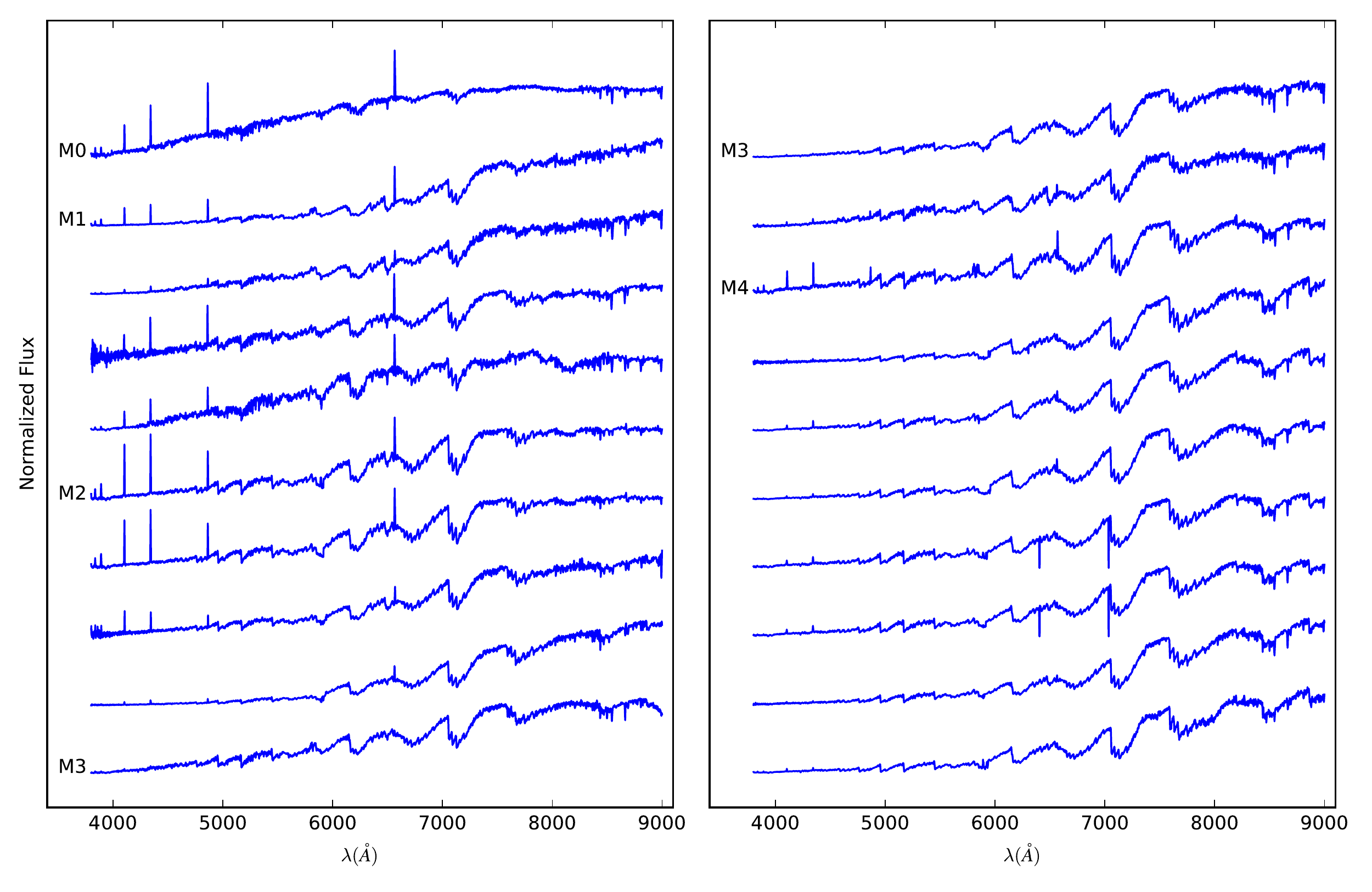}\\
\includegraphics [scale=0.67] {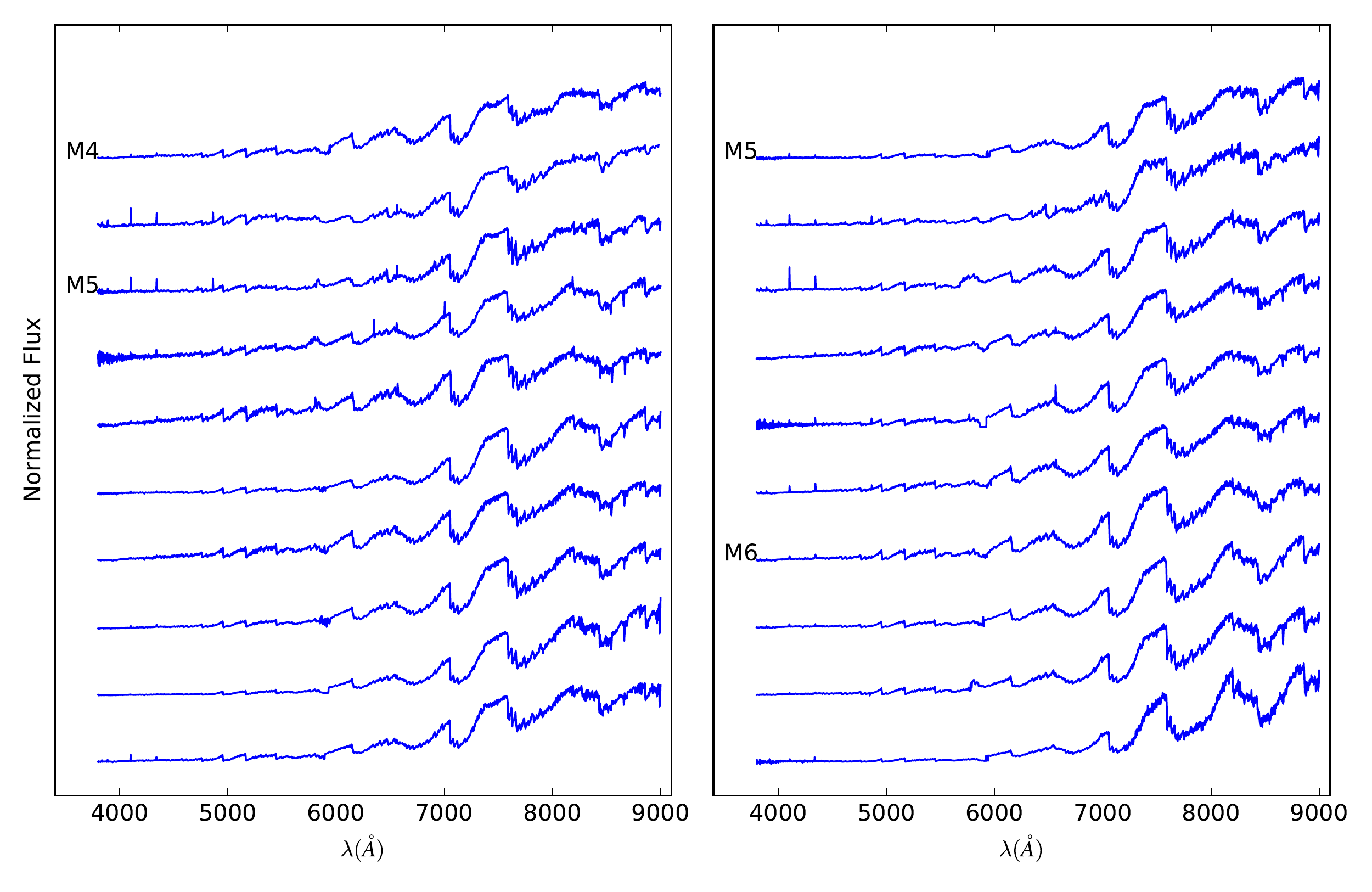}\\
\end{tabular}
}
\end{figure*}
\clearpage
\begin{figure*}
\center
\subfigure[]{%
\begin{tabular}{c}
\includegraphics [scale=0.67] {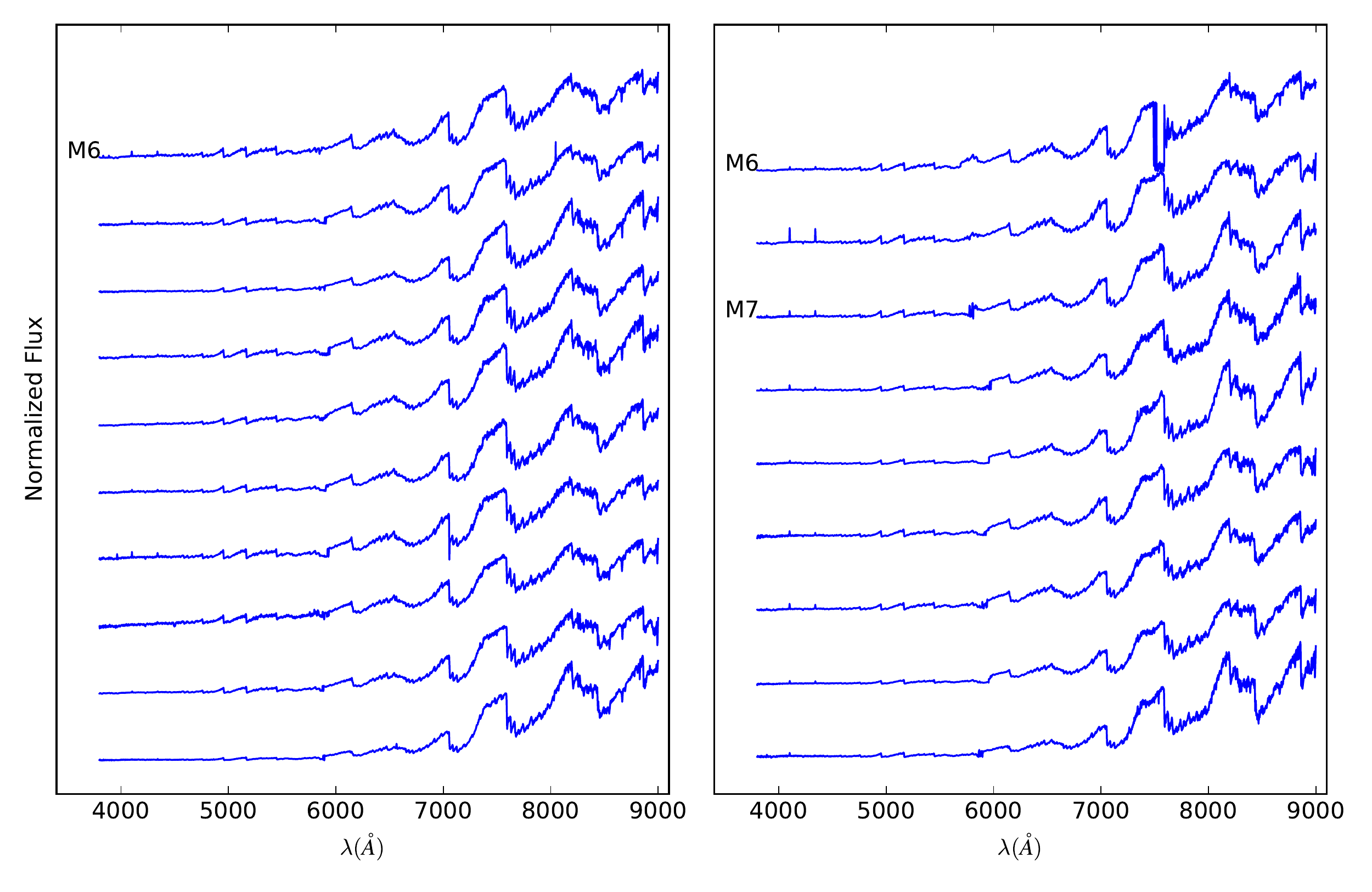}\\
\includegraphics [scale=0.67] {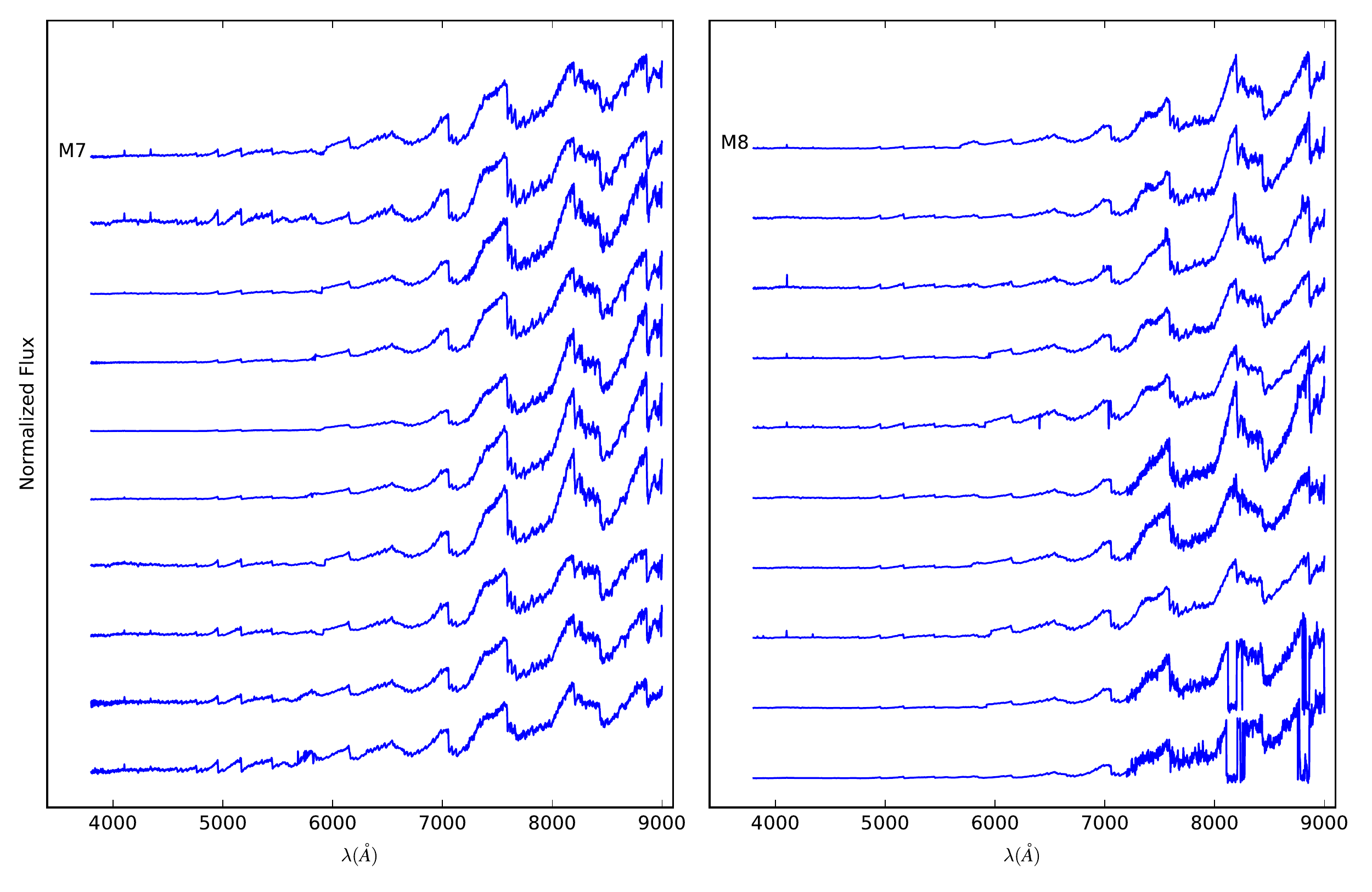}\\
\end{tabular}
}
\end{figure*}
\clearpage
\begin{figure*}
\center
\subfigure[]{%
\begin{tabular}{c}
\includegraphics [scale=0.67] {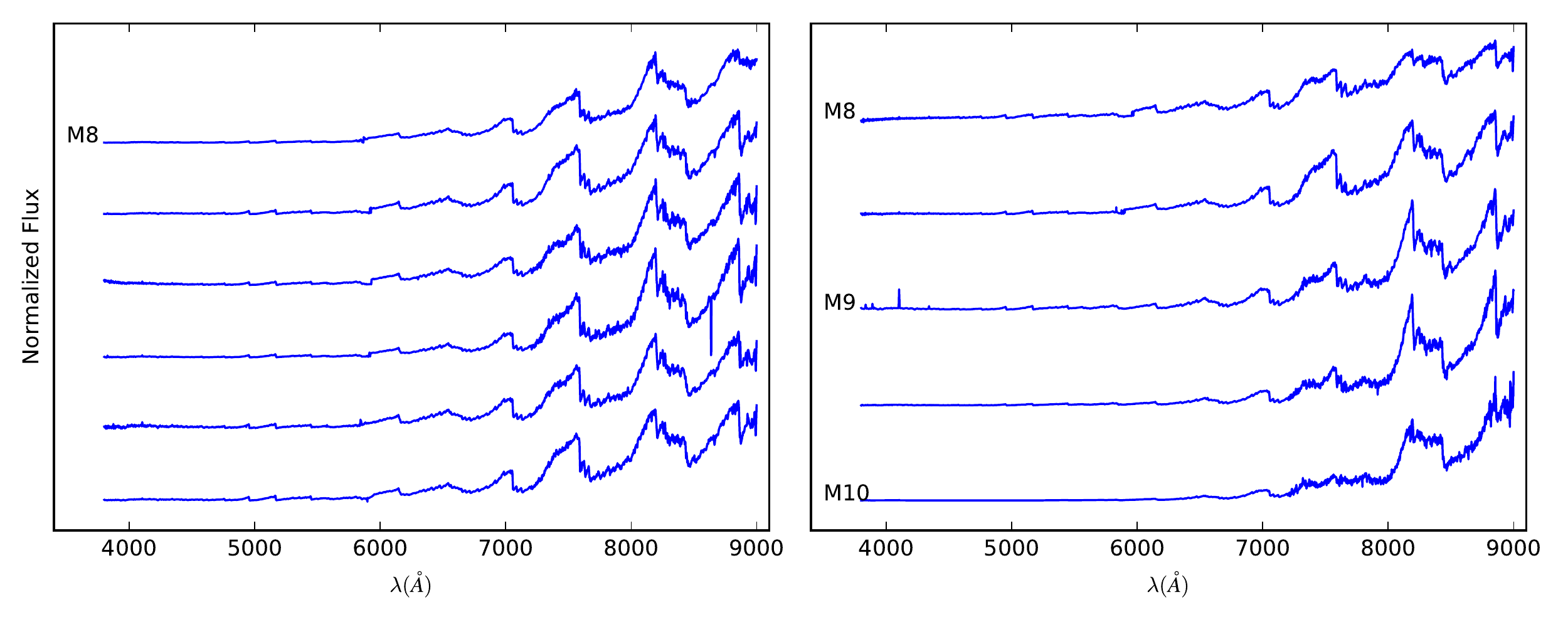}\\
\includegraphics [scale=0.67] {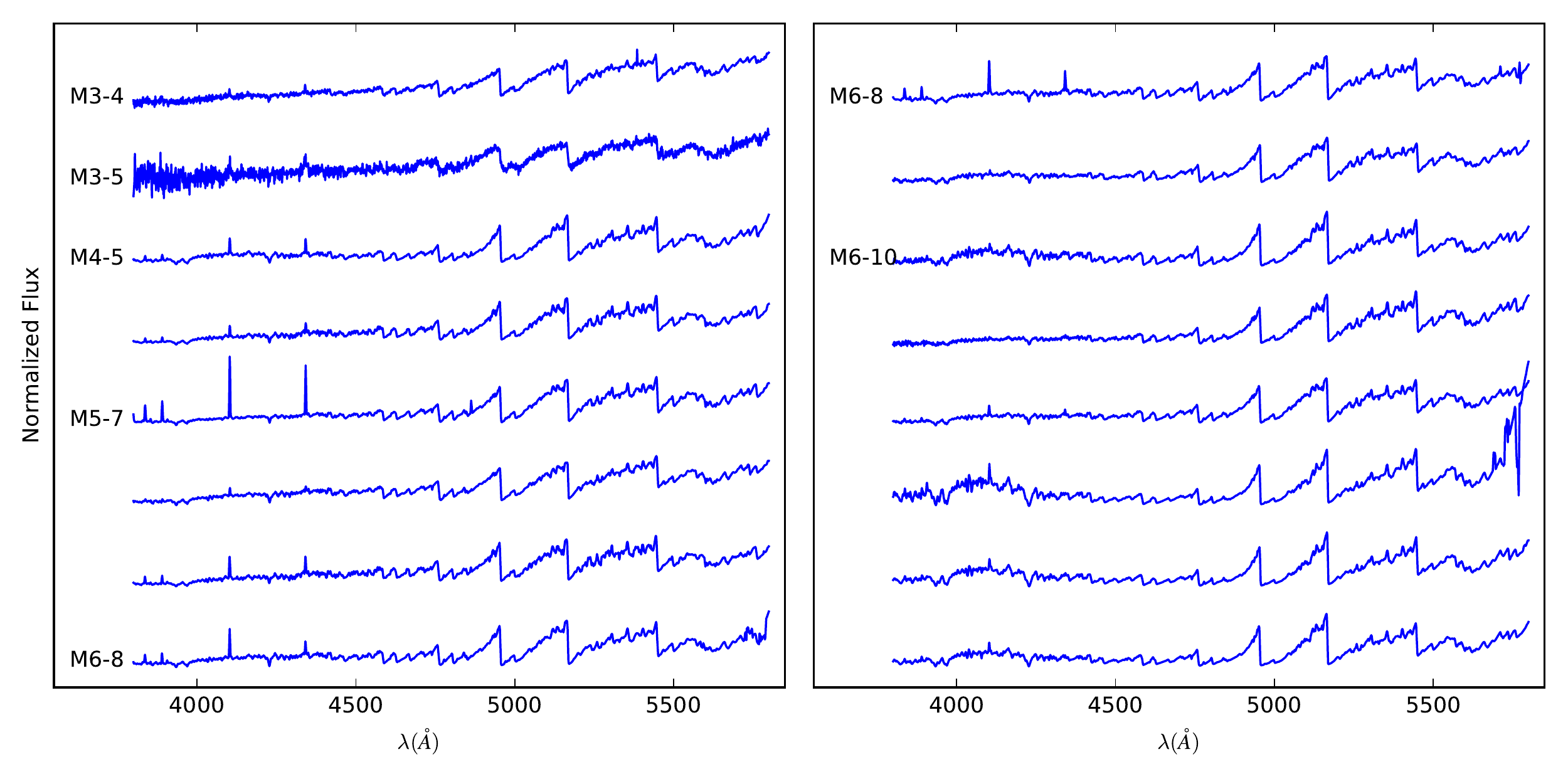}\\
\end{tabular}
}
\caption{O-rich Mira candidate spectra selected in Section
  \ref{subsec:by_eye} from LAMOST DR4. Temperature types of 16 of them
  can only be roughly determined because of long-wavelength
  saturation.
\label{fig:o_selected}}
\end{figure*}

\begin{figure*}
\center
\subfigure[]{%
\begin{tabular}{c}
\includegraphics [scale=0.67] {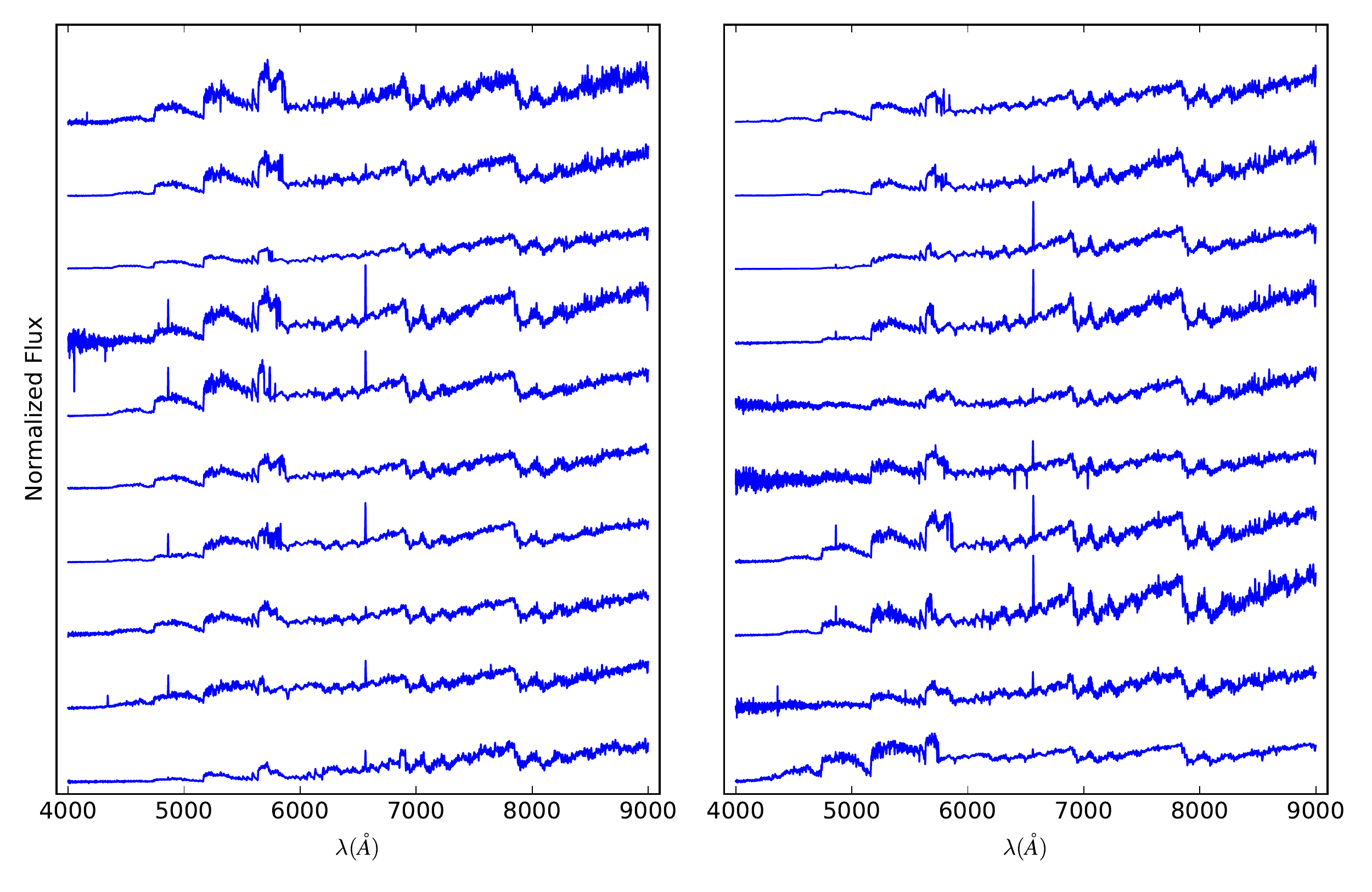}\\
\includegraphics [scale=0.67] {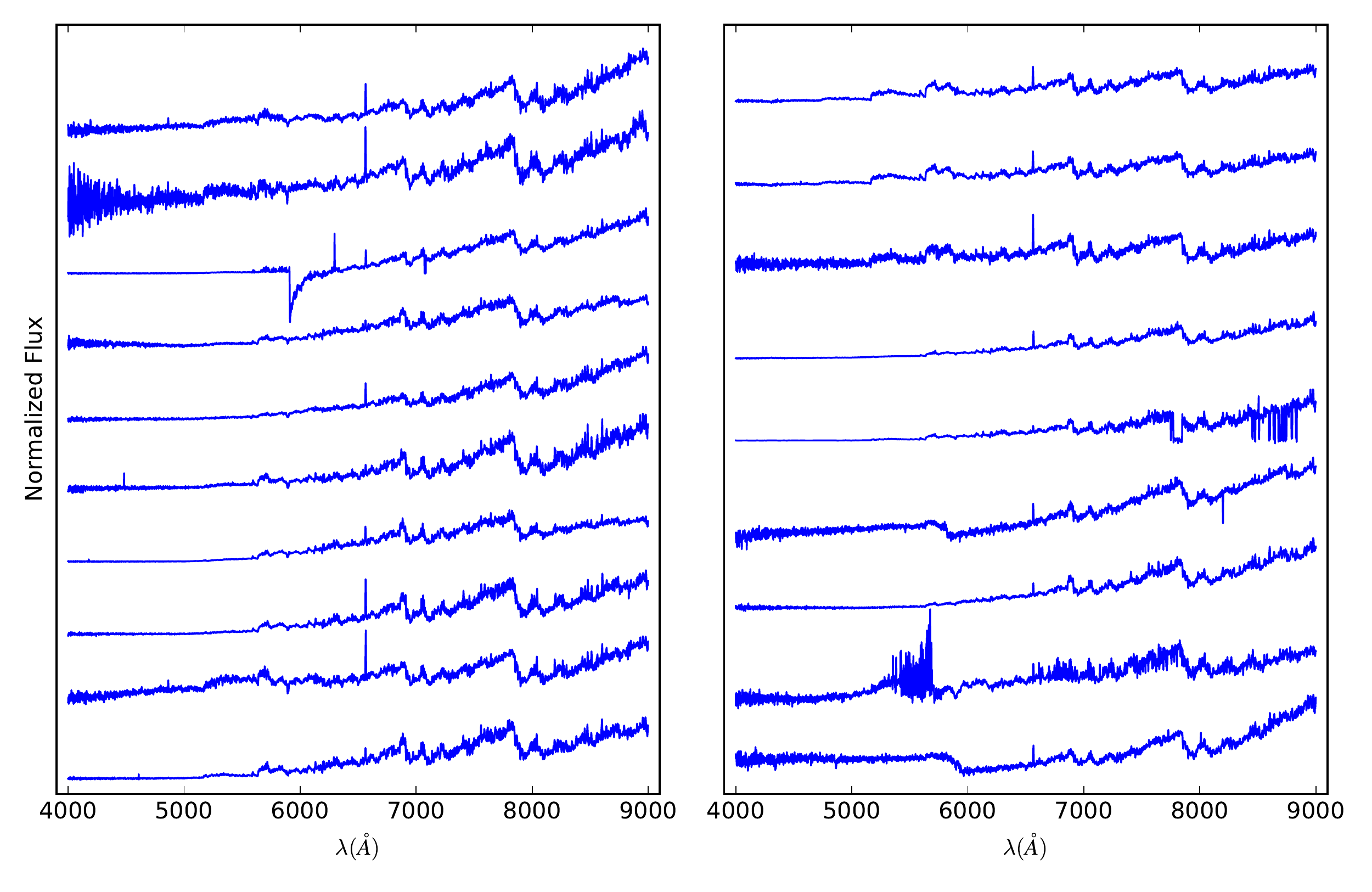}\\
\end{tabular}
}
\end{figure*}

\clearpage
\begin{figure*}
\center
\subfigure[]{%
\begin{tabular}{c}
\includegraphics [scale=0.67] {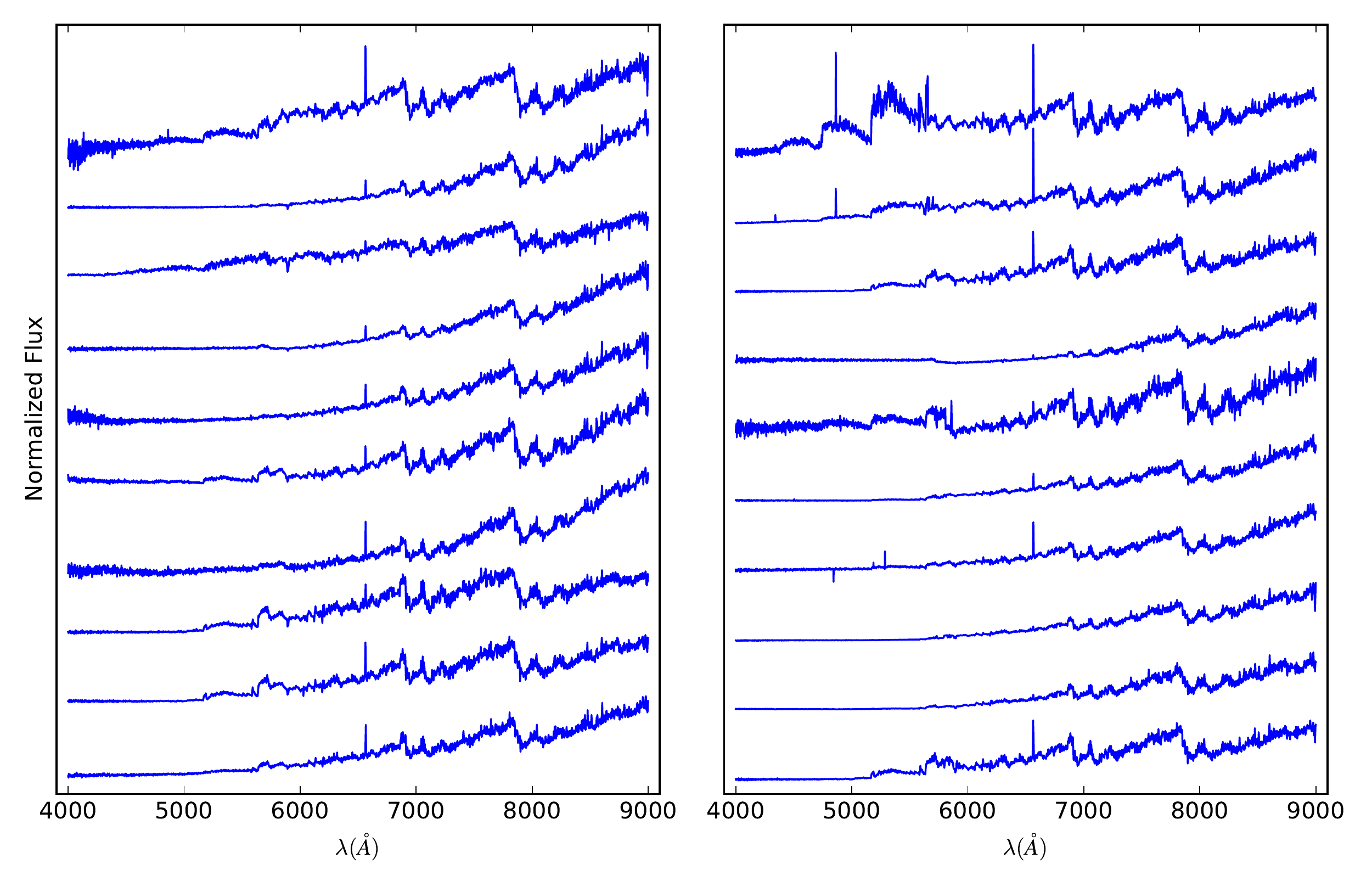}\\
\includegraphics [scale=0.67] {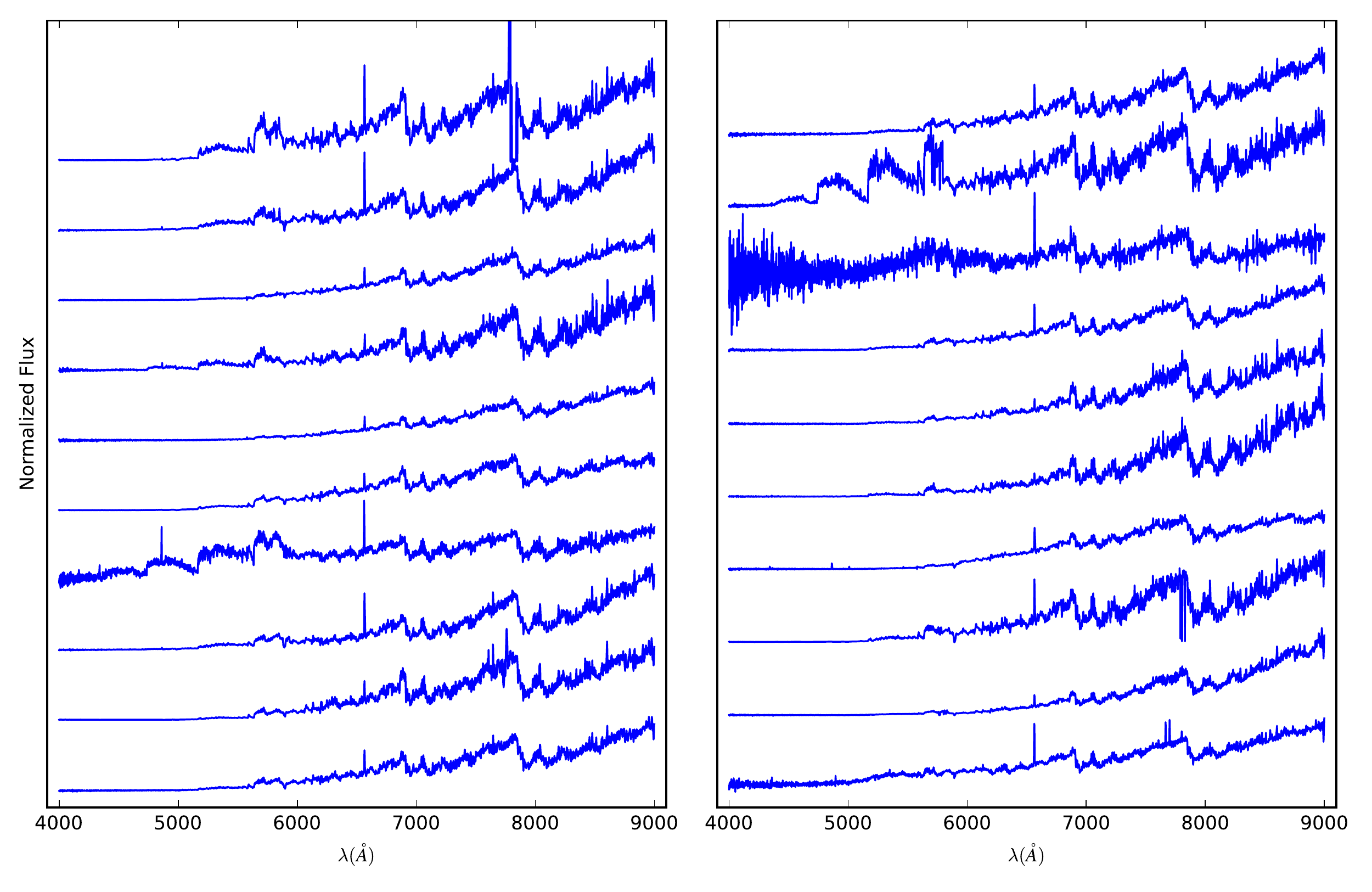}\\
\end{tabular}
}
\end{figure*}
\clearpage

\begin{figure*}
\center
\subfigure[]{%
\begin{tabular}{c}
\includegraphics [scale=0.67] {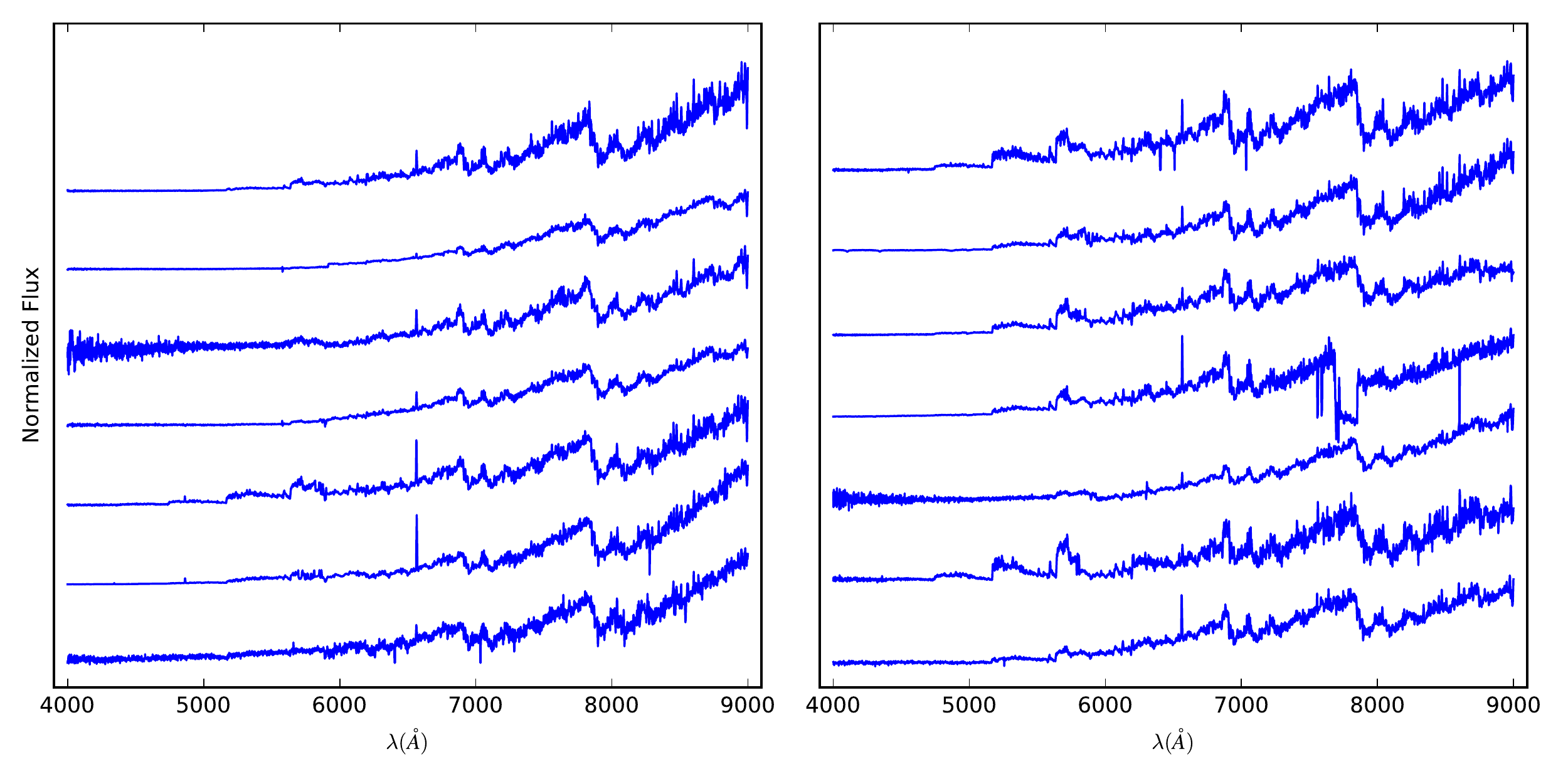}\\
\end{tabular}
}
\caption{C-rich Mira candidate spectra selected in Section
  \ref{subsec:by_eye} from LAMOST DR4.
\label{fig:c_selected}}
\end{figure*}
\end{document}